\documentclass[pra,a4paper,showpacs,superscriptaddress]{revtex4}
\usepackage{amsmath}
\usepackage{amsfonts}
\usepackage{graphicx}
\usepackage{longtable}
\newcommand{\be}{\begin{equation}}
\newcommand{\ee}{\end{equation}}

\newcommand{\ba}[1]{\left(\begin{array}{#1}}
\newcommand{\ea}{\end{array}\right)}
\begin{document}
\title{Majorana representation of symmetric multiqubit states} 
\author{A. R. Usha Devi}
\email{arutth@rediffmail.com}
\affiliation{Department of Physics, Bangalore University, 
Bangalore-560 056, India}
\affiliation{Inspire Institute Inc., Alexandria, 22303, USA.}
\author{Sudha}
\affiliation{Department of Physics, Kuvempu University, Shankaraghatta, Shimoga-577 451, India.}
\affiliation{DAMTP, Centre for Mathematical Sciences, Wilberforce Road, Cambridge, CB3 0WA, UK.}
\author{A. K. Rajagopal} 
\affiliation{Inspire Institute Inc., Alexandria, 22303, USA.}

\date{\today}

\begin{abstract} 
As early as 1932, Majorana had proposed that a pure permutation symmetric state of $N$ spin-$\frac{1}{2}$ particles can be represented by
N spinors, which correspond geometrically to $N$ points on the Bloch sphere. Several decades after its conception, the Majorana representation has recently attracted a great deal of attention in connection with multiparticle entanglement. A novel use of this representation led  to the classification of entanglement families of permutation symmetric qubits -- based on the number of distinct  spinors  and their arrangement in constituting the multiqubit state. An elegant approach to explore how correlation information of the {\em whole}  pure symmetric state gets imprinted in its {\em parts} is developed for specific entanglement classes of symmetric states. Moreover,  an elegant and simplified method to evaluate geometric measure of entanglement in $N$-qubit states obeying exchange symmetry has been developed based on the distribution of the constituent Majorana spionors over the unit sphere. Multiparticle entanglement being a key resource in several quantum information processing tasks, its deeper understanding is essential. In this review, we present a detailed description of the Majorana representation of pure symmetric states and its applicability in investigating various aspects of multiparticle entanglement.         
\end{abstract}
\maketitle

\renewcommand\thesection{\arabic{section}} 
\section{Introduction}
\label{intro}
Due to their experimental significance and mathematical elegance~\cite{sym1,sym1a,sym1b,sym2,sym3}, multiqubit states obeying exchange symmetry form an important class among quantum states. The class of symmetric states comprises of the well-known Greenberger-Horne-Zeilinger(GHZ)~\cite{ghz}, W, and Dicke states~\cite{dicke} etc. Mathematical simplicity in addressing 
$N$-qubit  states obeying permutation symmetry results because the states are confined to the $N+1$ dimensional subspace of the $2^N$ dimensional Hilbert space. The $N+1$ dimensional subspace is spanned by the Dicke states, $\{\vert N/2, N/2-l\rangle,  l=0,1,2,\ldots,N\}$, which are the simultaneous eigenstates of the squared collective angular momentum operator $J^2$ and its $z$-component $J_z$. An elegant geometrical representation for  multiqubit symmetric states in terms of $N$-points on the Bloch sphere $S^2$ was proposed by Majorana~\cite{majorana} as early as 1932. The representation of multiqubit states based on their characteristic $N$-qubits (spinors), the so-called {\em Majorana representation} (MR) for symmetric states~\cite{majorana,1945,penrose,makela} has been  immensely useful in diverse branches of physics~\cite{1945,jpa4,arxiv4,jpa5,ejtp6} in general and in quantum information science~\cite{solano, mixed,usa1,usa2,markham1,markham2,gebastin,markham3} in particular.  The significance of MR in characterizing entanglement in multiqubit symmetric states has been realized in recent years and the avenues appear to be expanding. While the SLOCC classification of symmetric states in terms of the distinct spinors characterizing the state has been accomplished using the MR~\cite{solano, mixed}, the reducibility/irreducibility features of multiparty correlations in several important classes of states could well be captured~\cite{usa1,usa2} using it. In fact, an ingenious use of this representation allows one to characterize how the multiparty correlation is imprinted in its parts for a class of non-symmetric states too, which is a generalized set of states related to a particular symmetric class~\cite{usa1}. Quantification of multiparty entanglement is another important aspect where the MR finds its applicability. Geometric measure of entanglement~\cite{Shimony, Wei} -- a useful measure of entanglement for multiqubit pure states -- has been realized to have a natural interpretation~\cite{markham1,markham2} in terms of the arrangement of $N$-spinors on the Bloch sphere, as given by MR. Identifying maximally entangled symmetric states for each $N$ has been possible utilizing this feature~\cite{markham1, markham2,gebastin, markham3} and this has paved way for arriving at some novel results while analyzing highly entangled states in the so-called platonic solids~\cite{markham3}. This review aims at capturing the essence of MR and how it has been put to use towards the understanding of multiqubit entanglement. 

The organization of the article is as under: Sec.~2 gives a detailed description of MR of symmetric multiqubit states, the Majorana spinors characterizing the state and their geometric representation. The SLOCC classification of symmetric states based on the number and arrangement of spinors constituting the state is given in Sec.~3.  We discuss the "whole and its parts" issue in the context of some specific SLOCC class of symmetric states. More specifically,  we show that not all states that are interconvertible into one another through SLOCC operations exhibit the same reducibility/irreducibility of correlations, with the help of an explicit three qubit example.  We also demonstrate that only two of the $N-1$ qubit reduced density matrices uniquely determine the family of $N$ qubit  pure symmetric states, comprised of two distinct Majorana spinors   and also another related class of non-symmetric states.  A brief review on geometric measure of entanglement and how the geometric representation of symmetric states given by Majorana  leads to a quantification of their entanglement, characterized through geometric measure, is given in Section~6. Section~7 contains a brief summary.    

\section{Majorana representation}
\label{majoranarep} 

 In his novel 1932 paper~\cite{majorana} (which had not received much attention at that time) Ettore Majorana proposed that a pure spin $j=\frac{N}{2}$ quantum state can essentially be represented as a {\em symmetrized} combination of $N$ constituent spinors as follows:  
\begin{equation}
\label{Maj}
\vert \Psi_{\rm sym}\rangle={\cal N}\, \sum_{P}\, \hat{P}\, \{\vert \epsilon_1, \epsilon_2, 
\ldots  \epsilon_N \rangle\}, 
\end{equation} 
where 
\begin{equation}
\label{spinor}
\vert\epsilon_l\rangle= 
\cos(\beta_l/2)\, e^{-i\alpha_l/2}\, \vert 0\rangle +
\sin(\beta_l/2)\, e^{i\alpha_l/2} \, \vert 1\rangle,\ \ l=0,1,2,\ldots, N,
\end{equation}
denote the spinors constituting the symmetric state $\vert \Psi_{\rm sym}\rangle$;
 $\hat{P}$ corresponds to the set of all $N!$ 
permutations of the spinors (qubits) and ${\cal N}$ corresponds to an overall normalization factor. 

For example, two and three qubit symmetric pure states have the following representations in terms of 
the Majorana spinors:
\begin{eqnarray}
\vert \Psi^{(2)}_{\rm sym}\rangle&=&{\cal N}\,  \left[\vert \epsilon_1, 
 \epsilon_2\rangle+  \vert \epsilon_2,\epsilon_1\rangle\right] \\
\vert \Psi^{(3)}_{\rm sym}\rangle&=&{\cal N}\,\left[  \vert \epsilon_1,
 \epsilon_2, \epsilon_3\rangle+  \vert \epsilon_3,\epsilon_1, \epsilon_2\rangle  + \vert \epsilon_2, 
 \epsilon_3, \epsilon_1\rangle \right. \nonumber \\ 
&& \left.  +
 \vert \epsilon_2,\epsilon_1,\epsilon_3\rangle+
 \vert \epsilon_3,\epsilon_2,\epsilon_1\rangle+ 
 \vert \epsilon_1, \epsilon_3, \epsilon_2\rangle \right].
\end{eqnarray}
Eq.(\ref{Maj}) corresponds to the {\em Majorana representation} of an arbitrary symmetric state $\vert \Psi_{\rm sym}\rangle$ of $N$ qubits in terms of the constituent spinors $\vert \epsilon_l \rangle,\  l=1,2,\ldots N$.

On the other hand, states of $N$-qubits obeying exchange symmetry get restricted to 
a $(N+1)$ dimensional Hilbert space spanned by the collective basis vectors $\left\{\left\vert \frac{N}{2}, l-\frac{N}{2}\right\rangle, l=0,1,2,
\ldots N \right\}$ where,    
\begin{equation}
\label{Dicke} 
\left\vert \frac{N}{2}, l-\frac{N}{2}\right\rangle =\frac{1}{\sqrt{^N C_l}}\,[\vert \underbrace{0, 0, \ldots}_{l}, 
\underbrace{1, 1, \ldots}_{N-l}\rangle +\ {\rm Permutations}\ ] 
\end{equation}
are the $N+1$ Dicke  states -- expressed in the standard qubit basis $\vert 0\rangle,\ \vert 1\rangle$ and  
 $^N C_l=\frac{N!}{l!\,(N-l)!}$ denotes the binomial coefficient.
An arbitrary pure symmetric state of $N$ qubits obeying exchange symmetry may thus be expressed as,    
\begin{equation}
\label{sympure1} 
\vert \Psi_{\rm sym}\rangle =\sum_{l=0}^{N}\, c_l\, \left\vert \frac{N}{2}, l-\frac{N}{2}\right\rangle,
\end{equation}
and is completely specified by the $(N+1)$ complex coefficients $c_l.$  Eliminating an overall phase and normalizing  the 
state  implies that $N$ complex parameters are required to 
completely characterize a pure symmetric state (\ref{sympure1}) of $N$ qubits.

While (\ref{sympure1}) offers a suitable parametrization of the symmetric multiqubit system in terms of the the collective parameters $c_l$, the MR (\ref{Maj}) leads to an intrinsic geometric picture of the  system in terms of $N$-points on the unit sphere $S^2$. (Note that the spinors $\vert \epsilon_l\rangle$ $l=1,2,\ldots, N$ of (\ref{spinor}) correspond geometrically to $N$ points on the so-called {\em Majorana sphere} $S^2$~\cite{makela,jpa4,markham1,markham2,gebastin, markham3}) -- with the pair of angles $(\alpha_l,\beta_l)$ determining the orientation of each point on the sphere).

The equivalence between the parameters $c_l$ of the collective representation (\ref{sympure1}) and that of  the MR (\ref{Maj}) can be established in an elegant manner~\cite{usa1} as detailed in the following. 
\begin{enumerate}
%\item Let us first note that the $2N$ orientation angles $(\alpha_l,\beta_l)$, $l=1,2,\ldots,N$, specifying the spinors (\ref{spinor}) of (\ref{Maj}) %may be conveniently  parametrized  in terms of the  $N$ complex parameters $z_l$ as,  
%\begin{equation}
%\label{z}
%z_l=\tan\frac{\beta_l}{2}e^{i\alpha_l},\ \ l=1,2,\ldots,N
%\end{equation}  
%and   the spinors $\vert\epsilon_l\rangle$ of (\ref{spinor}) may thus be expressed as 
%\begin{equation}
%\label{spinor2}
%\vert \epsilon_l\rangle=A_l\, \left[\vert 0 \rangle + z_l\,\vert 1 \rangle\right], \ \ A_l=e^{-i\alpha_l}\,\cos(\beta_l/2).
%\end{equation} 

\item A symmetric pure state is transformed into another symmetric pure state under identical rotations 
$R\otimes R\otimes \ldots \otimes R$ on all the spinors of (\ref{Maj}) (which corresponds  to an equivalent collective rotation  
${\cal R}$ on the state (\ref{sympure1}) in the $(N+1)$ dimensional symmetric sub-space). 
\item Under identical rotation through $R^{-1}(\alpha_s,\beta_s,0)\otimes R^{-1}(\alpha_s,\beta_s,0)\otimes 
\ldots  $, where $\alpha_s, \beta_s$ correspond to the orientation of any one of the spinors in (\ref{Maj}), it may 
be identified that 
\begin{equation}
\label{rl}
\langle 1,1\ldots , 1\vert R^{-1}(\alpha_s,\beta_s,0)\otimes R^{-1}(\alpha_s,\beta_s,0)\otimes 
\ldots   \vert \Psi_{\rm sym}\rangle\equiv 0.
\end{equation}
This is because the rotation $R_s^{-1}\otimes R_s^{-1}\ldots \otimes R_s^{-1}$ takes one of the spinors $\vert \epsilon_s\rangle$ with orientation angles $(\alpha_s, \beta_s)$ to $\vert 0\rangle$ i.e., it aligns the spinor $\vert \epsilon_s\rangle$ in the positive $z$-direction. 
Then, every   term in the  superposition (\ref{Maj}) of the rotated state has atleast one 
$\vert 0\rangle$ and so, the projection $~\langle 1,1,~\ldots,~ 1~\vert~ 
R_s^{-1}\otimes R_s^{-1}\otimes 
\ldots R_s^{-1}\vert \Psi^{(N)}_{\rm Sym}\rangle$ of the rotated state in the `all-down' direction
 vanishes. 
\item Eq.~(\ref{rl}) holds good for  collective rotations ${\cal R}_s^{-1}=R^{-1}_s\otimes 
R^{-1}_s\otimes\ldots\otimes 
R^{-1}_s,\ \ s=1,2,\ldots, N,$ which orient {\em any} one of the constituent  
spinors $\vert\epsilon_s\rangle$ in the positive $z$-direction. In other words, there exist $N$ rotations 
$R^{-1}_s=R^{-1}(\alpha_s,\beta_s,0), s=1,2,\ldots , N$ -- in general -- which lead to the same result  (\ref{rl}). 
  
\item In terms of the alternate representation (\ref{sympure1}) of the symmetric state $\vert \Psi_{\rm sym}\rangle$, (\ref{rl}) leads to 
\begin{eqnarray}
\label{prelim}
\left\langle \frac{N}{2},-\frac{N}{2}\left\vert 
R^{-1}(\alpha_s,\beta_s,0)\otimes R^{-1}(\alpha_s,\beta_s,0)\otimes 
\ldots  \otimes R^{-1}(\alpha_s,\beta_s,0) \right \vert \Psi_{\rm sym}\right\rangle&=&0\nonumber \\ 
\Rightarrow\ \ \left\langle \frac{N}{2},-\frac{N}{2}\left \vert {\cal R}_s^{-1}(\alpha_s,\beta_s,0) \left\{\sum_{l=0}^N\, c_l\, \right \vert \frac{N}{2},l-\frac{N}{2}\right \rangle\right\}&=&0 \nonumber \\
{\rm i.e.,} \ \ \ \sum_{l=0}^N\, c_l\, D^{N/2*}_{l-N/2,-N/2}(\alpha_s,\beta_s,0)=0, &&
\end{eqnarray}
where we have denoted 
$R^{-1}(\alpha_s,\beta_s,0)\otimes R^{-1}(\alpha_s,\beta_s,0)\otimes 
\ldots  \otimes R^{-1}(\alpha_s,\beta_s,0)~=~{\cal R}^{-1}_s(\alpha_s,\beta_s,0)$ in the collective $(N+1)$ dimensional symmetric subspace of $N$ qubits and 
$$D^{N/2\dag}_{-N/2,\, l-N/2}=\langle N/2,-N/2\vert{\cal R}^{-1}_l 
 \vert N/2,l-N/2\rangle,$$
  represents the collective rotation   
 in the Wigner-$D$ representation~\cite{Rose}. 
Substituting the explicit form of the $D$-matrix~\cite{Rose}, i.e.,
\begin{equation}
\label{dmatrix}
\left[D^{N/2\dag}(\alpha,\beta,0)\right]_{-N/2,\, l-N/2}=D^{N/2*}_{l-N/2,-N/2}(\alpha,\beta,0)=\sqrt{^N\, C_l}\, 
\left[\cos\left(\frac{\beta}{2}\right)\right]^{N-l}\, \left[-\sin\left(\frac{\beta}{2}\right)\right]^{l} \, e^{i(l-\frac{N}{2})\alpha},
\end{equation}  
and on subsequent simplification we obtain,
\begin{eqnarray}
\label{cmj02}
{\cal A}\, \sum_{l=0}^N (-1)^l\,\sqrt{^N\, C_l}\,  c_l\,  \, z^{l}&=&0\,  
\end{eqnarray} 
where $z=\tan\left(\frac{\beta}{2}\right)\,e^{i\, \alpha}$ and 
the overall coefficient ${\cal A}=\cos^N\left(\frac{\beta}{2}\right)\,e^{-i\frac{N\,\alpha}{2}}$. 

In other words, 
the $N$ roots $z_l=\tan\left(\frac{\beta_l}{2}\right)\,e^{i\, \alpha_l}, l=1,2,\ldots N$ of the Majorana polynomial $P(z)$
\begin{equation} 
\label{Mp}
P(z)=\sum_{l=0}^N\, (-1)^l\, \sqrt{^N\, C_l}\,  c_l\,  \, z^{l}
\end{equation}
 determine the orientations $(\alpha_l,\beta_l)$ of the  spinors 
constituting the  $N$-qubit symmetric state, in terms of the collective parameters $c_l$.
\end{enumerate}

It may be noted that the orientations of all the $N$ constituent spinors may not be determined in the cases where the Majorana Polynomial $P(z)$ is  of degree $r<N$ (i.e., when some of the coefficients $c_l,\ \  r< l\leq N$  are zero). To see this, let us consider the example of Dicke states (\ref{Dicke}). We have only one of the coefficients non-zero i.e., $c_l=\delta_{l,r}$.  The corresponding Majorana polynomial $P(z)$ reduces to  $P(z)=(-1)^r\, \sqrt{^N\, C_r}\, z^{r}$. The $r$-fold degenerate root  $z=0$ of the polynomial leads to the specification of the spinor orientation angles $\beta_l=0,\ \ \alpha_l=$arbitrary, $l=1,2,\ldots, r$ -- leading to the identification $\vert\epsilon_l\rangle\sim \vert 0\rangle,\ l=1,2,\ldots r$ (up to an overall phase) of the constituent spinors. There is no further information about the remaining $N-r$ spinors constituting the state in terms of the Majorana Polynomial (\ref{Mp}).  It is convenient to recast the  polynomial in terms of $z'=\frac{1}{z}=\cot \left(\frac{\beta_l}{2}\right)\,e^{-i\alpha_l}$ and following the same procedure outlined above, we obtain 
\begin{eqnarray}
\label{cmj03}
{\cal A'}\  \sum_{l=0}^N (-1)^l\,\sqrt{^N\, C_l}\,  c_{\frac{N}{2}-l}\,  \, z'^{N-l}&=&0,  
\end{eqnarray} 
where  ${\cal A'}=\sin^N\left(\frac{\beta_l}{2}\right)\,e^{i\alpha_l\, N/2}$. 

We thus obtain, 
\begin{equation} 
\label{Mp2}
P(z')=\sum_{l=0}^N\, (-1)^{N-l}\, \sqrt{^N\, C_l}\,  c_{\frac{N}{2}-l}\,  \, z'^{N-l}.
\end{equation}
The $N-r$ roots of the  polynomial (\ref{Mp2}) determine the orientations of the remaining $N-r$ spinors $\vert\epsilon_l\rangle$, $l=r+1,\,r+2,\,\ldots N$, constituting the state $\vert \Psi_{\rm sym}\rangle$. In particular, for Dicke states $\left\vert \frac{N}{2},r-\frac{N}{2}\right\rangle$, it is easy to see that (\ref{Mp2}) leads to $(N-r)$-fold degenerate root $z'=0$ which in turn corresponds to 
$\beta_l=\pi/2,\ \alpha_l=$ arbitrary i.e., $\vert \epsilon_l \rangle \equiv \vert 1 \rangle$, $l=r+1,\,r+2,\, \ldots N$. 
It may be readily seen that except for the  {\em all-up} ({\em all down}) $N$-qubit Dicke states $\left\vert \frac{N}{2}, \frac{N}{2}\right\rangle\equiv \vert 0,\,0,\ldots,0\rangle$   ( $\left\vert \frac{N}{2},-\frac{N}{2}=\right\rangle\equiv \vert 1,\,1,\ldots,1 \rangle$), for which the Majorana Polynomial  $P(z)=(-1)^N\, z^N$ ($\ P(z')=(-1)^N z'^{N}$) results in $N$-fold degenerate root,  
 all the other  Dicke states $\left\vert \frac{N}{2},l-\frac{N}{2}\right\rangle,\ l\neq 0, N$ are characterized by two distinct spinors, $\vert 0\rangle$, $\vert 1 \rangle$ each occurring $r$ and $N-r$ times respectively. 

The   GHZ state $\frac{1}{\sqrt{2}}\left[\left\vert \frac{N}{2},\frac{N}{2}\right\rangle+ 
\left\vert \frac{N}{2},-\frac{N}{2}\right\rangle\right]$ of $N$ qubits  satisfy the polynomial equation 
$1 +(-1)^N\, z^{N}=0$, solutions of which are  $N^{\rm th}$ roots of unity (when $N$ is odd)
$z_{l}=e^{\frac{2\pi\, i\, l}{N}};\ l=0,1,2,\ldots N-1$ (when $N$=even, we have
$z_{l}=e^{\frac{2\pi\, i}{N}(l-\frac{1}{2})}$). The associated  Majorana spinors are given by,  
$\vert \epsilon_{l}\rangle=\sqrt{\frac{z_l}{2}}\left[ \vert 0\rangle + z_l \vert 1\rangle\right]$, $l~=~0,1,2,\ldots, N~-~1.$ 

We list some examples of   symmetric states of two and three qubits and the corresponding constituent spinors in Table.~1. 
\begin{table}
{\scriptsize\begin{tabular}{|c c|c|c|c|}
\hline
 & $\begin{array}{cc} {\rm Symmetric\ state} \\ {\rm in\ the} \\ {\rm collective\ basis}\\ \left\vert
\frac{N}{2},l-\frac{N}{2}\right\rangle \end{array}$ & $\begin{array}{c}{\rm Polynomial\ equation}\\ {\rm and \ its \  %%@
solutions}\end{array}$ & Majorana spinors 
 & $\begin{array}{c} {\rm Symmetrization\ of}\\ {\rm Majorana\ spinors\ as\ in\ Eq.~(\ref{Maj})} \\ 
 ({\rm expressed\ in\ the}\\ {\rm standard\ qubit\ basis})\end{array}$
 \\
& &  & & \\
\hline 
 N=2 & $\frac{\vert 1,1\rangle + \vert 1,-1\rangle}{\sqrt{2}}$
& $\begin{array}{c} 1+ z^2=0, \\ 
z_{1,2}=e^{\pm i\frac{\pi}{2}}, \beta_{1,2}=\frac{\pi}{2}, \alpha_{1,2}=\pm\frac{\pi}{2}\end{array}$
& $\begin{array}{c} \vert \epsilon_{1}\rangle=
  \frac{e^{-i\frac{\pi}{4}}}{\sqrt{2}}\left(\vert0\rangle+  i \vert 1\rangle\right) \\
  \\ \vert \epsilon_{2}\rangle=
  \frac{e^{i\frac{\pi}{4}}}{\sqrt{2}}\left(\vert0\rangle-  i \vert 1\rangle\right)\end{array}$ & $\frac{\vert 0,0\rangle + \vert 1,1\rangle}{\sqrt{2}}$ \\
\hline 
  & $\frac{\vert 1,1\rangle - \vert 1,-1\rangle}{\sqrt{2}}$
& $\begin{array}{c} z^2-1=0,\\ z_{1,2}=\pm 1, \beta_{1,2}=\frac{\pi}{2}, \\ 
\alpha_{1,2}=0 \end{array}$ & $\begin{array}{c}\vert \epsilon_{1}\rangle=
  \frac{1}{\sqrt{2}}\, \left(\vert0\rangle+ \vert 1\rangle\right)  \\ 
   \vert \epsilon_{2}\rangle=
  \frac{1}{\sqrt{2}}\, \left(\vert0\rangle- \vert 1\rangle\right)
  \end{array}$  & $\frac{\vert 0,0\rangle - \vert 1,1\rangle}{\sqrt{2}}$\\        
\hline 
 & $\vert 1,0\rangle $
& $\begin{array}{c} z=0,\ \ z^{-1}=1  \\
z_{1}=0, z_2=z^{-1}=1 \\  \beta_{1,2}=0, \pi, \alpha_{1,2}={\rm arbitrary}\end{array}$
 & $\begin{array}{c}\vert \epsilon_{1}\rangle=\vert0\rangle \\ 
\vert \epsilon_{2}\rangle=
  \vert1\rangle \end{array}$ & $\frac{\vert 0,1\rangle + \vert 1,0\rangle}{\sqrt{2}}$\\        
\hline 
N=3  &  $\frac{\left\vert \frac{3}{2},\frac{3}{2}\right\rangle
+\left\vert \frac{3}{2},-\frac{3}{2}\right\rangle}{\sqrt{2}}$
& $\begin{array}{c} 1- z^3=0, \\
z_{r}=e^{\frac{2\pi\, i\, r}{3}},   \\ 
\beta_{r}=\frac{\pi}{2}, \alpha_{r}=\frac{2\pi\, r}{3}, r=0,1,2.\end{array}$
& $\begin{array}{c}\vert \epsilon_{r}\rangle=
  \sqrt{\frac{z_r}{2}}\, \left(\vert0\rangle+ z_r\, \vert 1\rangle\right),  \\ \ r=0,1,2\end{array}$
& $\frac{\vert 0,0,0\rangle+\vert 1,1,1\rangle}{2}$ \\
\hline 
  &  $\frac{\left\vert \frac{3}{2},\frac{3}{2}\right\rangle
-\left\vert \frac{3}{2},-\frac{3}{2}\right\rangle}{\sqrt{2}}$
& $\begin{array}{c}z^3+1=0, \\ 
z_{r}=e^{\frac{2\pi\, i\,}{3}(r-\frac{1}{2})} \\
\beta_{r}=\frac{\pi}{2}, \alpha_{r}=\frac{2\pi\, }{3}(r-\frac{1}{2}), r=0,1,2 \end{array}$
& 
$\begin{array}{c}\vert \epsilon_{r}\rangle=
  \sqrt{\frac{z_r}{2}}\, \left(\vert0\rangle+z_r\, \vert 1\rangle\right),  \\ \ r=0,1,2\end{array}$
  & $\frac{\vert 0,0,0\rangle - \vert 1,1,1\rangle}{2}$
  \\ 
\hline
&  $\frac{\left\vert \frac{3}{2},\frac{1}{2}\right\rangle
\pm \left\vert \frac{3}{2},-\frac{1}{2}\right\rangle}{\sqrt{2}}$
& $\begin{array}{c} z^2\mp z=0\\  
z_{1}=\pm 1, z_2=0, z_3^{-1}=0;\\
\beta_{1}=\pm \frac{\pi}{2},\, \alpha_1=\mp\pi; \beta_{2}=0; \beta_3=\pi.\end{array}$
& $\begin{array}{c}\vert \epsilon_{1}\rangle=
  \frac{1}{\sqrt{2}}\left(\vert0\rangle\pm \vert 1\rangle\right),  
 \\  \vert \epsilon_{2}\rangle=\vert 0\rangle, \ 
 \vert \epsilon_{3}\rangle=\vert 1\rangle\end{array} $ 
 & $\begin{array}{l}
\frac{1}{\sqrt{6}}\, [\vert 0,0,1\rangle+\vert 0,1,0\rangle \\ 
+\vert 0,0,1\rangle 
 \pm \vert 0,1,1\rangle\\ 
 \pm \vert 1,0,1\rangle\pm \vert 1,1,0\rangle]
 \end{array}$  \\       
  \hline
 &  $\left\vert \frac{3}{2}, -\frac{1}{2}\right\rangle$
& $\begin{array}{c} z=0, z^{-2}=0 \\
z_{1}=0,z_2^{-1}=z_3^{-1}=0,   \beta_{1}=0, \beta_{2,3}=\pi \\ 
\alpha_{1,2,3}={\rm abitrary} \end{array}$
& $\vert \epsilon_{1}\rangle=\vert0\rangle, \ \vert \epsilon_{2,3}\rangle=\vert 1\rangle$ 
 & $\frac{1}{\sqrt{3}}\, [\vert 0,0,1\rangle+\vert 0,1,0\rangle  
+\vert 0,0,1\rangle]$
 \\
 \hline        

\end{tabular}}
\caption{ Majorana spinors for some two and three qubit symmetric states}
\end{table}

\section{Entanglement classification of multiqubit symmetric states} 
\label{classification}
Multiparticle entanglement  can be of different kinds~\cite{Dur}. Two multiparty states have the same kind of entanglement if they
can be obtained from each other other via stochastic local operations
and classical communication (SLOCC) with nonzero probability. It is well-known that three qubit GHZ and W states are inequivalent under SLOCC and are representatives of inequivalent three party entanglement. Understanding  inequivalent classes of multiparticle entanglement, which are not interconvertible into each other under SLOCC operations is of fundamental importance~\cite{Dur,Ver,Lamata,solano}. It has been identified that the number of inequivalent multiparticle entanglement classes grows rapidly with the increase of the number of parties~\cite{Dur,Lamata, solano}. This poses increasing algebraic complexity in the identification of inequivalent entanglement classes as the  the number of parties increase. However, when one restricts to the set of permutation symmetric multiqubit states, the MR, discussed in Sec.~2, offers an elegant approach towards the  SLOCC classification of  entanglement families -- based entirely on the number and arrangement of the independent spinors (qubits) constituting the pure symmetric multiqubit state~\cite{solano}. More recently, innovative experimental schemes  have been proposed to generate a large variety of symmetric multiqubit photonic states~\cite{newexpt1,newexpt2}. In the following, we outline the approach of Bastin et.al~\cite{solano} in identifying the SLOCC classification of symmetric multiqubit pure states based on the MR.

\subsection{SLOCC classification of symmetric multiqubit pure states}  

Any two $N$-party {\em pure}  states $\vert\phi\rangle$, $\vert \psi\rangle$  are interconvertible, with  non-zero probability of success, 
by means of SLOCC if and only if there exists an invertible local operation (ILO)~\cite{Dur} ${\bf A}_1\otimes {\bf A}_2\otimes \ldots \otimes{\bf A}_N$ such that $\vert \phi\rangle=({\bf A}_1\otimes {\bf A}_2\otimes \ldots \otimes {\bf A}_N)\vert\psi\rangle.$  
Restricting ourselves to the set of permutation symmetric multiqubit states,  it  suffices to consider  transformations of the form ${\bf A}^{\otimes N}={\bf A}\otimes {\bf A}\otimes\ldots \otimes {\bf A}$, comprising only {\em identical} ILOs on all the qubits to verify the SLOCC equivalence~\cite{solano,bastin}. This identification is significant in that MR~\cite{majorana} of symmetric states offers itself to recognize how different entanglement families emerge.

As we have noted in Sec.~2,  the roots of the Majorana polynomial (\ref{Mp}) (and (\ref{Mp2})) could be degenerate  and hence not all the  $N$ constituent spinors of a pure symmetric $N$ qubit state are distinct. Let $\vert \epsilon_1\rangle, \vert \epsilon_2\rangle,\ldots, \vert \epsilon_d\rangle,\ d\leq N$ be the number of distinct spinors, in a $N$ qubit pure symmetric state (\ref{Maj}).  Then, the list of numbers $$\left\{n_1,\,n_2,\ldots ,n_d;\ \ n_1\geq n_2\geq\ldots \geq n_d; \ \ \  n_1+n_2+\ldots+n_d=N\right\}$$ 
 corresponds respectively to the number of times the independent spinors $\vert\epsilon_i\rangle$, ($i=1,2,\ldots d\leq N$) appear in the symmetric state (\ref{Maj}) under consideration. The number $d\leq N$, called the {\em diversity degree} and the list of numbers $\left\{n_1,\,n_2,\ldots ,n_d;\  n_1\geq 
n_2\geq \ldots n_d,\  \sum_{i=1}^{d} 
n_i=N\right\}$, called the {\em degeneracy configuration}, form the key elements in the classification of  pure symmetric states~\cite{solano}. The different classes  (based on the number of distinct spinors and their arrangement in a given $N$-qubit symmetric state)  are denoted by $\{{\cal D}_{n_1,n_2,\ldots n_d}\}$. 
 An identical ILO ${\bf A}^{\otimes N}$ transforms a symmetric state belonging to the class $\{ {\cal D}_{n_1,\,n_2,\ldots, n_d}\}$
  to another state of the {\em same} class. More explicitly, we have 
\begin{eqnarray}
\vert D_{n_1,n_2\ldots, n_d}\rangle \ \stackrel{\rm  ILO}{\longrightarrow}\ \vert D'_{n_1,\,n_2\ldots, n_d}\rangle
={\bf A}^{\otimes N}\,\vert D_{n_1,n_2\ldots, n_d}\rangle
\end{eqnarray}  
with the constituent spinors transforming as 
$\vert\epsilon'_i\rangle~=~A\, \vert\epsilon_i\rangle$, $i=1,2,\ldots d$. This forms the main basis of the SLOCC classification of symmetric pure 
states~\cite{solano}. 

\begin{enumerate}
\item $\{{\cal D}_N\}$: When all the $N$ solutions of the Majorana polynomial are identically equal, the corresponding class of  symmetric states is given by 
\begin{equation}
\vert D_N\rangle=\vert \epsilon,\epsilon,\ldots \epsilon\rangle,
\end{equation} 
where the diversity degree  $d=1$; the states belonging to this family of 
separable symmetric states  is denoted by $\vert D_N \rangle$.  

\item $\{{\cal D}_{n_1,n_2};\ \  n_1=N-k, n_2=k=1,2,\ldots, [N/2]\}$: The  states with two distinct spinors have the form, 
\begin{equation}
\label{n1n2}
\vert D_{N-k,k}\rangle={\cal N}\,[\vert \underbrace{\epsilon_1, \epsilon_1,\ldots 
\epsilon_1}_{N-k}, \underbrace{\epsilon_2, \epsilon_2,\ldots \epsilon_2}_{k}\rangle+{\rm \, Permutations\,}]
\end{equation} 
where $k=1,2, \ldots [N/2]$.  

Dicke states $\left\vert \frac{N}{2}, k-\frac{N}{2}\right\rangle$ are the representative states of the entanglement class 
$\{{\cal D}_{N-k,k}\}$ with two independent spinors  and clearly, they 
are all inequivalent under  SLOCC (as the degeneracy classification  is different for each $k=1,2, \ldots [N/2]$).     

\item $\{{\cal D}_{1,1,\ldots, 1}\}$:  When the $N$ roots of the Majorana Polynomial (\ref{Maj}) are all distinct,
the pure symmetric states constitute  the class $\{ {\cal D}_{1,1,1,\ldots, 1}\}$ with diversity degree  $d=N$. Clearly, the $N$ qubit GHZ state is a representative of this entanglement class. 
\end{enumerate}

The number of SLOCC classes  grows  with the increase in the number of qubits: For $N=2$, there are only $2$ entanglement families given by 
$D_{2}$ (the separable class) and $D_{1,\,1}$; for $N=3$ there are $3$ SLOCC classes given by $D_{3}$, $D_{2,\,1}$ and $D_{1,\,1,\,1} $ etc.  In general, the number of entanglement families for a symmetric $N$-qubit state grows~\cite{solano}, based entirely on the partition of the number $N$ in the arrangement $\left\{n_1,\,n_2,\ldots ,n_d;\ n_1\geq n_2\geq\ldots\geq n_d;\   \sum_{i=1}^{d} 
n_i=N\right\}$. However, the Majorana classes  with diversity degree $d\geq 4$ contain a continuous range of SLOCC classes, depending on a continuous  parameter and the states with different value of this continuous  parameter  are not SLOCC convertible into each other~\cite{solano}. 
More recently~\cite{mixed} Bastin et. al have also extended the entanglement classification scheme for mixed symmetric multiqubit systems, based on the hierarchical families of different SLOCC classes, successively embedded into each other.

\section{Determining the whole pure state from its parts}

One among the basic issues of interest in quantum information theory is to learn about how much of the whole quantum state can be known from 
 its subsystems~\cite{Niel,SP1,SP2,SP2b,WL1,WL2,PS1,PS2}. The importance of knowing if higher order correlations in a multipartite system follow entirely from lower order ones involving few parties  has  been of interest in  many body physics~\cite{Coleman}. Construction of the many electron state with the knowledge of its two particle reduced density matrices has been discussed in a series of papers~\cite{M,M2,M3,M4,M5}. While it has been shown by Linden et.al~\cite{Linden,Linden2} that $N$-party entanglement cannot, in general, be reversibly transformed into entanglement of two parties, Linden, Popescu and Wootters~\cite{SP1, SP2}  proved a striking result that reduced states of a smaller fraction of the parties specify most of the {\em generic} multiparty pure states {\em uniquely}. Walck and Lyons~\cite{WL1, WL2} showed that the $N$ party GHZ  states and their local unitary equivalents are the only exceptions to this result and the correlations in a multi-qubit GHZ state are {\em irreducible}. Preeti Parashar and Swapan Rana have shown that $N$ qubit W class states can be uniquely determined by their bipartite marginals~\cite{PS1}.  Generalized Dicke-class states is another class of symmetric as well as non-symmetric states that is shown to possess reducible correlations~\cite{PS2}. In this section we discuss determining the whole pure symmetric $N$ qubit state of a specific SLOCC class from its $N-1$ party reduced states~\cite{usa1,usa2}.

\subsection{Irreducibility features of three qubit symmetric states of the class $\{{\cal D}_{1,1,1} \}$ with all distinct Majorana spinors:}

While the equivalence of quantum states under SLOCC is known to indicate that states belonging to the same equivalence class can be used to implement 
similar quantum information tasks~\cite{Dur}, here, we address the question    
"Do SLOCC interconvertible states possess similar  irreducibility features?" by considering specific examples of three qubit symmetric states belonging to the same SLOCC class $\{{\cal D}_{1,1,1}\}$.   With the help of this example, we demonstrate explicitly that the states of the same Majorana class could exhibit contrasting irreducibility features. GHZ state and its local unitary equivalent states are the only ones of the Majorana class $\{{\cal D}_{1,1,1}\}$,  which are undetermined by their two qubit reduced systems~\cite{SP1,WL1,WL2}.

We consider two specific examples of the SLOCC family $\{{\cal D}_{1,1,1}\}$,  the first being the three qubit 
GHZ state,  
\begin{equation}
\label{GHZ}
\vert {\rm GHZ}\rangle= \frac{1}{\sqrt{2}}\, [\vert 0,0,0\rangle +  \vert 1,1,1\rangle]. 
\end{equation}      
The Majorana polynomial equation (\ref{Maj}) for this state has a simple structure $1-z^3=0$, solutions of 
which are cube roots of unity $\omega, \omega^2,\omega^3=1$ and the corresponding spinors constituting the state %%@
are readily identified to be 
$\vert \epsilon_1\rangle =\frac{1}{\sqrt{2}}[\vert 0\rangle+\omega\, \vert 1\rangle]$,
$\vert \epsilon_2\rangle =\frac{1}{\sqrt{2}}[\vert 0\rangle+\omega^2\, \vert 1\rangle]$,%\nonumber \\
$\vert \epsilon_3\rangle =\frac{1}{\sqrt{2}}[\vert 0\rangle+ \vert 1\rangle].$ 
 GHZ state is fragile under the loss of a qubit, with vanishing pairwise concurrence~\cite{Wot,Wot2} for any pairs of 
two qubit reduced density matrices; but it exhibits genuine three-party entanglement~\cite{Dur,RaRe} with the 
maximum tangle~\cite{Kun}  $\tau=1$.  The state exhibits irreducible three party correlations which can not be  
determined by its reduced states~\cite{SP1,WL1, WL2}. 

We  consider another three qubit state which belong to the same SLOCC class $\{{\cal D}_{1,1,1}\}$ of three distinct Majorana spinors: 
\begin{equation}
\label{Wsup}
\vert \eta\rangle = \frac{1}{\sqrt{2}} [\vert {\rm W}\rangle +  \vert{\rm \bar{W}}\rangle]. 
\end{equation}         
This  is a superposition of the three qubit W state $\vert{\rm W}\rangle=\frac{1}{\sqrt{3}}[\vert 0,0,1\rangle+\vert 0,1,0\rangle+\vert 1,0,0\rangle]$  
 and its obverse state 
$\vert{\rm \bar{W}}\rangle=\frac{1}{\sqrt{3}}[\vert 1,1,0\rangle+\vert 1,0,1\rangle+\vert 
0,1,1\rangle].$ 
The state $\vert \eta\rangle$ has genuine three party entanglement, quantified in terms of the tangle 
$\tau=1/3$, and it is also robust under the loss of qubits -- as reflected through the concurrence $C=1/3$ for 
any pairs of two qubits.  The three qubit symmetric state $\vert \eta\rangle$ given by (\ref{Wsup}) satisfies the Majorana polynomial 
equation $z(z-1)=0$  and the corresponding spinors constituting the state are 
$\vert \epsilon'_1\rangle =\vert 1\rangle,$\ 
$\vert \epsilon'_2\rangle =\frac{1}{\sqrt{2}}[\vert 0\rangle+ \vert 1\rangle],$ 
$\vert \epsilon'_3\rangle = \vert 0\rangle.$

The states $\vert {\rm GHZ}\rangle$ and the W superposition 
state $\vert \eta\rangle$   can be {\em locally} converted from one another,  with the help of an identical ILO i.e.,
 $\vert {\rm GHZ}\rangle=A\otimes A\otimes A\, \vert \eta\rangle$,  where 
$A=\left(\begin{array}{cc} 1 & \omega \\ 1 & \omega^2\, \end{array}\right).$
The corresponding Majorana spinors  of the states $\vert \eta\rangle$ and $\vert{\rm GHZ}\rangle$ are  related to each other 
up to an overall factor:   
$A\, \vert\epsilon_1'\rangle=\sqrt{2}\omega\, \vert\epsilon_1\rangle,$   
$A\, \vert\epsilon_2'\rangle=-\omega^2\, \vert\epsilon_2\rangle,$ 
and $A\, \vert\epsilon_3'\rangle=   \sqrt{2}\,  \vert\epsilon_3\rangle$.  
 
We now explicitly show that the higher order correlation in  the W superposition state $\vert \eta\rangle$ is imprinted in its two qubit reduced states. We follow the approach of Linden et.al.,~\cite{SP1} in demonstrating this feature of the three qubit W superposition state.   

Let us suppose that a mixed  three qubit state $\gamma$ too has the same two-qubit reduced system $\varrho_{12}$, 
as that of the W superposition state $\vert \eta\rangle.$ Denoting the pure state $\vert \Gamma\rangle$ to be  
containing the three qubits and the environment such that  
$${\rm Tr}_{E}[\vert\Gamma\rangle\langle \Gamma\vert]=\gamma,$$ 
the two party reduced state $\varrho_{12}$ can be expressed as  
$$\varrho_{12}={\rm Tr}_{3, E}[\vert \Gamma\rangle\langle \Gamma\vert].$$ 
The two qubit reduced system $\varrho_{12}$ of the pure  
state $\vert\eta\rangle$ is a rank-2 state given by,  
\begin{eqnarray}
\label{ws2}
\varrho_{12}&=&\vert \chi_0\rangle\langle \chi_0\vert + 
\vert \chi_1\rangle\langle \chi_1\vert, 
\end{eqnarray}
where
\begin{eqnarray*}
\vert \chi_0\rangle&=&\frac{1}{\sqrt{6}}[|1,0\rangle+|0,1\rangle+|1,1\rangle],\\ 
   \vert \chi_1\rangle&=&\frac{1}{\sqrt{6}}[|0,0\rangle+|0,1\rangle+|1,0\rangle]. 
   \end{eqnarray*}
Given that the two party reduced state $\varrho_{12}$ also belongs to the extended pure state 
$\vert\Gamma\rangle$ (and in turn to the mixed state $\gamma$) of the three qubits and the environment, we must have  
\begin{eqnarray}
\label{ws3}
\vert\Gamma\rangle&=&\vert \chi_0\rangle\vert E_0\rangle +\vert \chi_1\rangle\vert E_1\rangle, \\  
\label{ws3'}
&&\langle E_i\vert E_j\rangle=\delta_{i,j},
 \end{eqnarray}     
In terms of the basis states of qubit 3, the states of the environment 
$\vert E_{0}\rangle, \vert E_{1}\rangle$ are given by
\begin{eqnarray}
\vert E_0\rangle&=&\vert 0\rangle\,  \vert e_{00}\rangle+\vert 1\rangle\,  \vert e_{01}\rangle, \nonumber\\ 
\vert E_1\rangle&=&\vert 0\rangle\,  \vert e_{10}\rangle+\vert 1\rangle\,  \vert e_{11}\rangle.
\end{eqnarray}   
Thus,   (\ref{ws3}) takes the following  form: 
\begin{eqnarray} 
\label{Gamma}
|\Gamma\rangle&=&\frac{1}{\sqrt{6}}[(|1_1,0_2,0_3\rangle+|0_1,1_2,0_3\rangle+|1_1,1_2,0_3\rangle)|e_{00}\rangle 
+(|1,0,1\rangle+|0,1,1\rangle+|1,1,1\rangle)|e_{01}\rangle \nonumber \\
&+& (|0,0,0\rangle+|0,1,0\rangle+|1,0,0\rangle)|e_{10}\rangle
+(|0,0,1\rangle +|0,1,1\rangle+|1,0,1\rangle)|e_{11}\rangle] 
\end{eqnarray}
Now, demanding that the reduced system $\varrho_{13}$ of $\vert \eta\rangle$ is also shared by  $\vert 
\Gamma\rangle$ leads to further constraints.  
\begin{enumerate}
\item First we compare  $\langle 0,1\vert \varrho_{13}\vert 0,1\rangle$,  from the states 
(\ref{Wsup}) and (\ref{ws3}): We have,  
$\langle 0,1|{\rm Tr}_{2}\,[\vert\eta\rangle \langle\eta\vert]\, |0,1\rangle= 
\frac{1}{3}$ and \break 
$\langle 0,1|{\rm Tr}_{2,E}\,[\vert\Gamma\rangle \langle\Gamma\vert]\, |0,1\rangle= 
\frac{1}{6}\langle e_{01}|e_{01}\rangle+\frac{1}{3}\langle e_{11}|e_{11}\rangle$ 
leading to  
$\langle e_{01}|e_{01}\rangle+2\langle e_{11}|e_{11}\rangle=2.$

\item Next, we compare  $\langle 1,1|\varrho_{13}|1,1\rangle$ evaluated from the states 
$\vert\eta\rangle$ and $\vert\Gamma\rangle$:  
We get, 
$\langle 1,1|{\rm Tr}_{2, E}\, [|\Gamma\rangle \langle\Gamma|]\,\vert 1,1\rangle=
\frac{1}{3}\langle e_{01}|e_{01}\rangle+\frac{1}{6}\langle e_{11}|e_{11}\rangle$ 
and $\langle 1,1|{\rm Tr}_{2}\, [|\eta\rangle \langle\eta|\vert\, |1,1\rangle=\frac{1}{6}$ 
implying,  
$2\langle e_{01}|e_{01}\rangle+\langle e_{11}|e_{11}\rangle=1.$ 
From these relations we obtain $\langle e_{11}|e_{11}\rangle=1,$ \ 
$\langle e_{01}|e_{01}\rangle=0$ (or $\vert e_{01}\rangle\equiv 0$).
Further, from the orthonormality (\ref{ws3'}) 
it follows that $\langle e_{00}\vert e_{00}\rangle=1,$ and  
$\vert e_{10}\rangle\equiv 0.$ 

\item Finally, a comparison of the matrix elements    
$\langle 0,0|{\rm Tr}_{2,E}\,[\vert\Gamma\rangle \langle\Gamma\vert]\, |0,1\rangle= 
\frac{1}{6}\, \langle e_{00}|e_{11}\rangle$ and 
 $\langle 0,0|{\rm Tr}_{2}\,[\vert\eta\rangle \langle\eta\vert]\, |0,1\rangle~=~ 
\frac{1}{6}$ 
lead to $\langle e_{00}|e_{11}\rangle=1$ or $\vert e_{11}\rangle\equiv \vert e_{00}\rangle.$ 
Thus,  the extended pure state (\ref{ws3}) should take the form 
$\vert \Gamma\rangle\equiv \vert \eta\rangle \, \vert e_{00}\rangle.$ 
\end{enumerate}
In other words,  the three qubit pure state $\vert \eta\rangle$  is uniquely determined by its two-qubit reduced %%@
systems and is therefore, reducible. 

This illustrative example of three qubits supports ( with the help of an independent and non-trivial proof)  the already existing general result~\cite{WL1,WL2} that only the $N$ qubit GHZ state and its local unitarily equivalents which remain undetermined by their reduced density matrices. Moreover, this clearly projects out the contrasting   
irreducibility features of two SLOCC interconvertible states (\ref{GHZ}) and (\ref{Wsup}) of the {\em same}  
entanglement family $\{{\cal D}_{1,1,1}\}$.

\subsection{Determining $\{{\cal D}_{N-k,k}\}$ SLOCC class of pure symmetric $N$ qubit states  from its $N-1$ qubit reduced density matrices} 

While it has been proved~\cite{WL1} that except for the $N$ qubit GHZ states and their unitary equivalents all other pure states are uniquely determined by their $N-1$ party marginals, {\em how many} $N-1$ party marginals  are required to uniquely specify a given $N$-qubit state--not unitarily equivalent to GHZ state-was not known.  

Let us consider the example of  $4$-qubit pure states considered by Ref.~\cite{WL1}:  
$\vert\chi_1\rangle=\frac{1}{\sqrt{3}}\left(\vert 0000\rangle+ \vert 0001\rangle+\vert 1111\rangle   \right)$ and 
$\vert\chi_2\rangle=\frac{1}{\sqrt{3}}\left(\vert 0000\rangle+ \vert 0001\rangle-\vert 1111\rangle   \right)$. These two states  are not unitarily equivalent to the $4$-qubit GHZ state; they both share the same $3$-qubit reduced density matrices, when partial trace over first, second and third qubits are taken -- whereas the partial trace over $4^{\rm th}$ qubit leads to distinct $3$-qubit marginals. In other words, all the four  $3$-qubit reduced density matrices are required to uniquely specify each of them. Examples of $N$ qubit states determined by smaller numbers of $N-1$ qubit reduced density matrices are therefore of interest.  

It may be mentioned here that Preeti Parashar and Swapan Rana~\cite{PS2} focussed on identifying the class of states which can be uniquely determined by reduced density matrices of smaller than $N-1$ parties. The present authors~\cite{usa1}  showed that the $\{{\cal D}_{N-k,k}\}$ SLOCC class of pure symmetric $N$ qubit states containing two distinct Majorana spinors and another related class of  non-symmetric states can be determined with the help of {\em only two} of their $N-1$ qubit marginals. We outline the approach of Ref.~\cite{usa1} in the following.

Let us consider a representative  symmetric state (\ref{n1n2}) of the entanglement family 
$\{ {\cal D}_{N-k,k}\}$ 
\begin{eqnarray}
\label{dnk}
\vert D_{N-k, k}\rangle &=& {\cal N}\, \sum_{P}\, \hat{P}\,\{ \vert \underbrace{\epsilon_1, \epsilon_1,
\ldots , \epsilon_1}_{N-k};\ \underbrace{\epsilon_2, \epsilon_2,\ldots , \epsilon_2}_{k}\rangle\}\nonumber \\
&=& {\cal N}\, R_1^{\otimes N}\, \sum_{P}\, \hat{P}\,\{ \vert \underbrace{0, 0,
\ldots , 0}_{N-k};\ \underbrace{\epsilon'_2, \epsilon'_2,\ldots , \epsilon'_2}_{k}\rangle\},
\end{eqnarray}
where $\epsilon_1=R_1\vert 0\rangle$ and $\epsilon_2=R_2\vert 0\rangle$, and 
\begin{equation}
\label{ep'}
\vert \epsilon'_2\rangle=R_1^{-1}R_2\vert 0\rangle=d_0\, \vert 0\rangle+d_1\, \vert 1\rangle,\ \ \vert d_0\vert^2+\vert d_1\vert^2=1,\ \ d_1\neq0.
\end{equation}
Substituting (\ref{ep'}) in (\ref{dnk}) and upon simplification, we obtain,
\begin{eqnarray}
\vert D_{N-k, k}\rangle &=& R_1^{\otimes N}\, \sum_{r=0}^k\,\sqrt{^N C_{r}}\, \alpha_{r}\, \left\vert\frac{N}{2},\frac{N}{2}-r \right\rangle,
\nonumber \\
\label{Ar}
{\rm where} \  \alpha_{r}&=&{\cal N}\,\, 
\frac{(N-r)!}{(N-k)! (k-r)!}\, d_0^{k-r}\, d_1^r. 
\end{eqnarray}
In other words, all symmetric states $\vert D_{N-k, k}\rangle$, constituted by two distinct Majorana spinors are equivalent 
(under local unitary transformations) to 
\begin{equation}
\label{nk'}
\vert D'_{N-k, k}\rangle=R_1^{-1\, \otimes N}\vert D_{N-k, k}\rangle=\sum_{r=0}^k\, \sqrt{^N C_{r}}\, \alpha_{r}\, \left\vert\frac{N}{2},\frac{N}{2}-r \right\rangle.
\end{equation}
As $d_1\neq 0$ (see Eq.~(\ref{ep'}), the coefficients  $\alpha_r$, ($r=0,\,1,\,2,\ldots,\,k$) are non-zero, except when $d_1=1, d_0=0$ -- in which case  the state  $\vert D'_{N-k, k}\rangle$ reduces to the Dicke state $\left\vert\frac{N}{2},\frac{N}{2}-k \right\rangle$ itself and then we have, 
$\alpha_r=\delta_{k,r}$.  We proceed to show that only two of the $N-1$ qubit reduced density matrices determine the whole state $\vert D'_{N-k, k}\rangle$, following an analogous procedure as in Ref~\cite{SP1}.   

We express the state $\vert D'_{N-k, k}\rangle$  in the qubit basis as 
\begin{eqnarray}
\label{dqubit}
\vert D'_{N-k,k}\rangle&=& \alpha_0\, |0_1,\,0_2,\,\cdots,\,0_N \rangle 
+\alpha_1 
\sum_{P}\hat{P}\{|1_1,0_2\,\cdots,\,0_{N-1},0_N \rangle\}+ 
 \alpha_2\, \sum_{P}\hat{P}\{|1_1,\,1_2,\,0_3\,\cdots,\,0_{N} 
\rangle\} +\cdots \cdots\nonumber \\ 
&&+\alpha_k\sum_{P}\hat{P}\{|1_1,\,1_2,\,\cdots,\,1_{k},\,0_{k+1},\cdots,\,0_N
\rangle\}=\vert \phi_0\rangle\, \vert 0_N\rangle+  \vert \phi_1\rangle\, \vert 1_N\rangle 
\end{eqnarray}
where
\begin{eqnarray}
\label{phidqubit1}
 \vert \phi_0\rangle&=&\alpha_0\vert 0_1,0_2,\cdots,0_{N-1} \rangle+\alpha_1\sum_{P} \hat{P}\{\vert 1_1,0_2,\cdots,0_{N-1} \rangle\} 
+ \alpha_2\sum_{P} \hat{P}\{|1_1,1_2,0_3,\ldots,0_{N-1} \rangle\} \nonumber \\
& &+\ldots+\alpha_k \sum_{P} \hat{P}\{|1_1,1_2,1_3,\cdots,1_{k},0_{k+1},\cdots,0_{N-1} \rangle\} \\
\vert \phi_1\rangle&=& \alpha_1\vert 0_1,0_2,\cdots,0_{N-1} \rangle+\alpha_2\sum_{P} \hat{P}\{\vert 1_1,0_2,\cdots,0_{N-1} \rangle\}
+\alpha_3 \sum_{P}\hat{P}\{|1_1,1_2,0_3,\ldots,0_{N-1} \rangle\}  \nonumber \\
&&+\ldots\ldots+\alpha_{k} \sum_{P} \hat{P}\{|1_1,1_2,1_3,\cdots,1_{k-1},0_{k},\cdots,0_{N-1} \rangle\}. 
\end{eqnarray} 

It is not difficult to see that the $N-1$ qubit reduced density matrix $\rho_{1,\,2,\ldots, N-1}$, obtained by tracing out the $N$th qubit from the state $\vert D'_{N-k, k}\rangle$, is a rank-2 mixed state  given by, 
\begin{eqnarray}
\label{nk'2}
\rho_{1,2,\ldots, N-1}&=&{\rm Tr}_{N}[\vert D_{N-k,k}'\rangle\langle 
D_{N-k,k}'\vert]=\vert \phi_0\rangle\langle \phi_0\vert + 
\vert \phi_1\rangle\langle \phi_1\vert. 
\end{eqnarray}

On supposing that a mixed  $N$ qubit state $\omega_N$ too shares the same  
$N-1$ qubit reduced system $\rho_{1,2,\ldots, N-1}$  one has  
\begin{eqnarray} 
\rho_{1,2,\ldots, N-1}&=&{\rm Tr}_{N}[\vert D_{N-k,k}'\rangle\langle 
D_{N-k,k}'\vert]={\rm Tr}_{N}[\omega_N] \nonumber \\ 
&=&\vert \phi_0\rangle\langle \phi_0\vert + 
\vert \phi_1\rangle\langle \phi_1\vert.  
\end{eqnarray} 
Considering the purification of the mixed state $\omega_N$, i.e., considering $\omega_N$ as a reduced system of 
an extended {\em pure} state $\vert\Omega_{NE}\rangle$ consisting of $N$ qubits and an environment $E$, one has 
\begin{equation}
{\rm 
Tr}_{E}[\vert\Omega_{NE}\rangle\langle \Omega_{NE}\vert]=\omega_N.
\end{equation} 
In order that the  pure state $\vert\Omega_{NE}\rangle$ (or the mixed state $\omega_N$) too shares the same $N-1$ qubit reduced density matrix $\rho_{1,2,\ldots, N-1}$, one must have
\begin{eqnarray}
\label{wp3}
\vert\Omega_{NE}\rangle&=&\vert \phi_0\rangle\vert E_0\rangle +\vert \phi_1\rangle\vert E_1\rangle, \\ 
\label{on} 
&& \langle E_i\vert E_j\rangle=\delta_{i,j}. 
\end{eqnarray}     
Here, the states $\vert E_{0}\rangle,\  \vert E_{1}\rangle$ are the ones containing the qubit labelled  $N$,  and the 
environment $E$. Expanding $\vert E_{0}\rangle,\  \vert E_{1}\rangle$ in  the basis states of the qubit $N$ as,   
\begin{eqnarray}   
\label{wp4}
\vert E_0\rangle&=&\vert 0_N\rangle\,  \vert e_{00}\rangle+\vert 1_N\rangle\,  \vert e_{01}\rangle \nonumber \\ 
\vert E_1\rangle&=&\vert 0_N\rangle\,  \vert e_{10}\rangle+\vert 1_N\rangle\,  \vert e_{11}\rangle,   
\end{eqnarray}
the state $\vert \Omega_{NE}\rangle$ can be re-expressed using (\ref{wp3}), (\ref{wp4}) as  
\begin{equation}
\label{omega}
 \vert\Omega_{NE}\rangle=\vert \phi_0\rangle \vert 0_N\rangle\,\vert e_{00}\rangle +\vert \phi_0\rangle \vert %%@
1_N\rangle\,\vert e_{01}\rangle 
+ \vert \phi_1\rangle \vert 0_N\rangle\,\vert e_{10}\rangle +\vert \phi_1\rangle \vert 1_N\rangle\,\vert %%@
e_{11}\rangle. 
\end{equation}
The states $\vert e_{ij}\rangle$, $i,\,j=0,\,1$ correspond to those  of the environment and they are not assumed to be orthonormal {\em apriori}.   

As both $\vert D_{N-k,k}'\rangle$ and $\vert\Omega_{NE}\rangle$ are sharing a common reduced density matrix 
$\rho_{1,2,\ldots, N-1}$, we wish to check the form of $\vert\Omega_{NE}\rangle$ so that it shares another reduced density matrix $\rho_{2,3,\ldots, N}$ of $\vert D_{N-k,k}'\rangle$. 

\begin{enumerate}
\item We first compare the matrix elements of 
$\rho_{2,\cdots,N}={\rm Tr}_1 \vert D'_{N-k,k} \rangle \langle D'_{N-k,k}\vert$   with  ${\rm Tr}_{1,E}[\vert \Omega_{NE}\rangle \langle\Omega_{NE}\vert]$ to find that 
\begin{eqnarray} 
\label{matrix1}
\langle 0_2,0_3\cdots,0_{N-k-1},1_{N-k},\cdots,1_{N}\vert{\rm Tr}_{1,E}[\vert \Omega_{NE}\rangle\langle 
\Omega_{NE}\vert]
\vert 0_2,0_3\cdots, 0_{N-k-1},1_{N-k},\cdots, 1_{N}\rangle &=&\vert\alpha_k\vert^2\langle e_{01}\vert e_{01}\rangle  \nonumber \\ 
\langle 0_2,0_3 \cdots, 0_{N-k-1},1_{N-k},\cdots, 1_{N}\vert{\rm Tr}_{1}[\vert D_{N-k,k}'\rangle\langle 
D_{N-k,k}'\vert]
\vert 0_2,0_3\cdots, 0_{N-k-1},1_{N-k},\cdots, 1_{N}\rangle&=&0   
\end{eqnarray} 
As $\alpha_k\neq 0$ for the states $\vert D_{N-k,k}'\rangle$, we must have 
$\vert e_{01}\rangle\equiv 0$. 
The simpler form of  $\vert\Omega_{NE}\rangle$  
obtained on putting $\vert e_{01}\rangle\equiv 0$ in (\ref{omega})  is given by 
\begin{equation}
\label{Omega2}
|\Omega_{NE}\rangle=|\phi_0\rangle|0_N\rangle |e_{00}\rangle+|\phi_1\rangle\left[|0_N\rangle|e_{10}\rangle+
|1_N\rangle|e_{11}\rangle\right]
\end{equation} 
On making use of the orthonormality relations $\langle E_{0}\vert E_{0}\rangle=1$, $\langle E_{0}\vert E_{1}\rangle=0$,    
we get
\begin{equation}
\label{ortho}
\langle e_{00}|e_{00}\rangle=1, \ \  \langle e_{00}|e_{10}\rangle=0.
\end{equation}

\item We now equate another matrix element of $\rho_{2,\cdots,N}$ obtained from both the states $\vert D_{N-k,k}'\rangle$ and $|\Omega_{NE}\rangle$:
\begin{eqnarray}
\label{matrix2}
&& \langle 0_2,0_3\cdots, 0_{N-k},1_{N-k+1},\cdots, 1_{N}\vert{\rm Tr}_{1,E}[\vert \Omega_{NE}\rangle\langle 
\Omega_{NE}\vert]
\vert 1_{2},\cdots, 1_{k+1},0_{k+2},0_3\cdots, 0_{N}\rangle
=\vert\alpha_k\vert^2 \langle e_{11}\vert e_{00}\rangle,\\
&&\langle 0_2,0_3\cdots, 0_{N-k},1_{N-k+1},\cdots, 1_{N}\vert{\rm Tr}_{1}[\vert D_{N-k,k}'\rangle\langle 
D_{N-k,k}'\vert]
\vert 1_{2},\cdots, 1_{k+1},0_{k+2},0_3\cdots, 0_{N}\rangle 
=\vert\alpha_k\vert^2.
\end{eqnarray} 
This leads to the identification,  $\langle e_{11}|e_{00}\rangle=1,$ as $\alpha_k\neq 0$. In view of (\ref{ortho}), we obtain 
$\vert e_{11}\rangle=\vert e_{00}\rangle+\vert e^{\perp}_{00}\rangle.$ Substituting this in the orthonormality condition 
$\langle E_1\vert E_1\rangle=1$, we readily obtain 
\begin{eqnarray*}
 \langle e_{10}\vert e_{10}\rangle + \langle e^\perp_{00}\vert e^\perp_{00}\rangle=0 \Rightarrow \vert e_{10}\rangle=0,\, \vert e^\perp_{00}\rangle=0
 \end{eqnarray*} 
which in turn implies that    
$$\vert e_{11}\rangle\equiv|e_{00}\rangle.$$

\end{enumerate}
Thus, we obtain  
\begin{eqnarray}
\label{Omega3A}
|\Omega_{NE}\rangle&=& \left(\vert \phi_0\rangle\,\vert 0_N \rangle+\vert \phi_1\rangle\,\vert 1_N \rangle\right)\vert e_{00}\rangle \nonumber \\
&=&\vert D'_{N-k,\,k}\rangle\, \vert e_{00}\rangle.
\end{eqnarray} 
This implies that the state 
$\vert D'_{N-k,\,k}\rangle$ is the  unique {\em whole} pure state that is consistent with 
its $N-1$ qubit reduced density matrices. We have employed only {\em two} of the $N-1$ reduced density matrices 
$\rho_{1,\,2,\cdots, N-1}$, $\rho_{2,\,3,\cdots, N}$ to arrive at this result. 
It may also be noted  here that any other choice of the second $N-1$ qubit reduced density matrix (obtained by tracing over {\em any} of one the qubits)  would have led us to the same result, though with appropriate choices of matrix elements in (\ref{matrix1}), (\ref{matrix2}).

\subsection{Unique determination  of a general class of non-symmetric $N$ qubit states through its parts}
The method illustrated for symmetric  states  $\vert D_{N-k,\,k}\rangle$  suggests a natural extension to a generalized family $D^G_{N-k,k}$ of  non-symmetric states. This family consists of states $\vert D^{G}_{N-k,k}\rangle$ which are a superpositions of the so-called generalized Dicke states $$\sum_{i}\,a^{(r)}_{i}\, \left[ |1_{P_{(i_1)}},1_{P_{(i_2)}},\ldots, 1_{P_{(i_r)}},0_{P_{(i_{r+1})}}\ldots ,0_{P_{(i_N)}} \rangle\right]$$  
obtained on associating an arbitrary coefficient $a^{(r)}_i$ with each term of the Dicke state (\ref{Dicke}). 

As we have shown in the previous section  that the  states $\vert D_{N-k,k}\rangle$ are local unitarily equivalent to the state 
\begin{eqnarray*}
\vert D'_{N-k,k}\rangle&=&\sum_{r=0}^k\, \sqrt{^N C_{r}}\, \alpha_{r}\, \left\vert\frac{N}{2},\frac{N}{2}-r \right\rangle, \\
\end{eqnarray*}
we construct the generalized non-symmetric $N$ qubit pure states from the state $\vert D'_{N-k,k}\rangle$ as follows:  
\begin{eqnarray}
\label{dgqubit}
 \vert D^{G}_{N-k,k}\rangle&=&\alpha_0\,a_0^{(0)}\, |0_1,0_2,\cdots,0_N \rangle+ \sum_{r=1}^k\alpha_r\left\{ \sum_{i=1}^{^N C_r}a^{(r)}_{i}\left[ |1_{P_{(i_1)}},1_{P_{(i_2)}},\ldots, 1_{P_{(i_r)}},0_{P_{(i_{r+1})}}\ldots,0_{P_{(i_N)}} \rangle\right]\right\} \nonumber \\
 &=&\alpha_0\, a_0^{(0)} |0_1,0_2,\cdots,0_N \rangle+\alpha_1
\left\{a^{(1)}_1|1_1,0_2,\cdots,0_N \rangle+a^{(1)}_2|0_1,1_2,\cdots,0_N \rangle +\cdots \cdots \right. \nonumber \\
&& + \left. a^{(1)}_N |0_1,0_2,\cdots,0_{N-1},1_N \rangle \right\} + \alpha_2\left\{a^{(2)}_1 |1_1,1_2,0_3,\cdots,0_N\rangle+a^{(2)}_2|1_1,0_2,1_3,0_4,\cdots,0_N \rangle \right. \nonumber \\
& & +\cdots + a^{(2)}_{\frac{(N-1)(N-2)}{2}}|0_1,\cdots,0_{N-3},1_{N-2},1_{N-1},0_N \rangle \nonumber \\ 
& & \left. + a^{(2)}_{\frac{(N-1)(N-2)}{2}+1}|1_1,0_2,0_3,\cdots,0_{N-1},1_N \rangle  +\cdots  + a^{(2)}_{\frac{N(N-1)}{2}}|0_1,0_2,\cdots,0_{N-2},1_{N-1},1_{N} \rangle\right\}+\cdots \nonumber \\ 
& & +\alpha_k\,\left\{\, a^{(k)}_1\, |1_1,1_2,\cdots,1_{k},0_{k+1},\cdots,0_N \rangle +\cdots + a^{(k)}_{^{N-1} C_k}\, |0_1,\cdots,0_{N-k-1},1_{N-k},1_{N-k+1},\cdots, 1_{N-1},0_N \rangle \right.  \nonumber \\
& & +\left.  a^{(k)}_{^{N-1} C_{k}+1}\, |1_1,1_2\cdots,1_{k-1}, 0_{k},\cdots,0_{N-1},1_N \rangle + \cdots +  a^{(k)}_{^{N}C_k}\, |0_{1},\cdots,0_{N-k},1_{N-k+1},\cdots, 1_N \rangle\right\}.   
\end{eqnarray}
Here $\alpha_r$'s are as given in (\ref{Ar}) and  
$\sum_{i}\,a^{(r)}_{i}\, \left[ |1_{P_{(i_1)}},1_{P_{(i_2)}},\ldots, 1_{P_{(i_r)}},0_{P_{(i_{r+1})}}\ldots ,0_{P_{(i_N)}} \rangle\right]$  
are the generalized Dicke states. We show that no  other (pure or mixed) $N$ qubit state, can share the same $N-1$ qubit subsystem density matrices as that of $\vert D^{G}_{N-k,k}\rangle$. The procedure adopted for this purpose is same as that employed in the subsection 3B.   

The state $\vert D^{G}_{N-k,k}\rangle$ can be expressed as  
\begin{eqnarray} 
\label{dgphi01}
\vert D^{G}_{N-k,k}\rangle&=&\vert \phi^{G}_0\rangle \vert 0\rangle_N + 
\vert \phi_1^{G}\rangle \vert 1\rangle_N, 
\end{eqnarray}
where 
\begin{eqnarray} 
\label{phig0}
 \vert \phi^{G}_0\rangle&=&\alpha_0\,a_0^{(0)} |0_1,0_2,\cdots,0_{N-1} \rangle+\sum_{r=1}^k \alpha_r\left\{ \sum_{i=1}^{^{N-1} C_r}a^{(r)}_{i} \left[ |1_{P_{(i_1)}},1_{P_{(i_2)}},\ldots, 1_{P_{(i_r)}},0_{P_{(i_{r+1})}}\ldots ,0_{P_{(i_{N-1})}} \rangle\right]\right\} \nonumber \\
 &=&\alpha_0\,a_0^{(0)} |0_1,0_2,\cdots,0_{N-1} \rangle+\alpha_1
\left\{a^{(1)}_1 |1_1,0_2,\cdots,0_{N-1}\rangle+ \cdots + 
a^{(1)}_{N-1}|0_1,\cdots,0_{N-2},1_{N-1} \rangle 
\right\} \nonumber \\
 &&+ \alpha_2 \left\{a^{(2)}_1 |1_1,1_2,0_3,\cdots,0_{N-1}\rangle+a^{(2)}_2 |1_1,0_2,1_3,0_4,\cdots,0_{N-1} \rangle
+\cdots  \right. \nonumber \\
 && \left.  +   a^{(2)}_{\frac{(N-1)(N-2)}{2}} |0_1,\cdots,0_{N-3},1_{N-2},1_{N-1}\rangle\right\}+\cdots\nonumber \\ 
&&    
+ \alpha_k\left\{ a^{(k)}_1 |1_1,1_2,\cdots,1_{k},0_{k+1},\cdots,0_{N-1} \rangle 
 +a^{(k)}_2 |1_1,1_2,\cdots,1_{k-1}, 0_{k},1_{k+1},0_{k+2},\cdots,0_{N-1}\rangle \right.  \nonumber \\ 
&& +
a^{(k)}_3 |1_1,1_2,\cdots,1_{k-2}, 0_{k-1},1_k,1_{k+1},0_{k+2}\cdots,0_{N-1}\rangle 
+\cdots  \nonumber \\ 
&& \left.  +a^{(k)}_k |0_1,1_2,\cdots,1_{k+1},0_{k+2}\cdots,0_{N-1}\rangle+  
 \cdots +a^{(k)}_{^{N-1} C_k}|0_1,\cdots,0_{N-k-1},1_{N-k},\cdots, 1_{N-1}\rangle\right\}, 
\end{eqnarray} 
and 
\begin{eqnarray} 
\label{phig1} 
 \vert \phi^{G}_1\rangle&=& \sum_{r=0}^{k-1} \alpha_{r+1}\left\{ \sum_{i=^{N-1} C_{r+1}+1}^{^{N} C_{r+1}}a^{(r+1)}_{i}\, \left[ |1_{P_{(i_1)}},1_{P_{(i_2)}},\ldots, 1_{P_{(i_{r})}},0_{P_{(i_{r+1})}}\ldots ,0_{P_{(i_{N-1})}} \rangle\right]\right\} \nonumber \\
&=&\alpha_1 a^{(1)}_N |0_1,0_2,\cdots,0_{N-1}\rangle
 +\alpha_2 \,\left\{ a^{(2)}_{\frac{(N-1)(N-2)}{2}+1} |1_1,0_2,0_3,\cdots,0_{N-1} \rangle+\cdots \right.\nonumber \\ 
 && + \left.
a^{(2)}_{\frac{N(N-1)}{2}} |0_1,0_2,\cdots,0_{N-2},1_{N-1} \rangle\right\} +\cdots+ \alpha_k\left\{  a^{(k)}_{^{N-1} C_k+1} |1_1,1_2\cdots,1_{k-1}, 0_{k},\cdots,0_{N-1} \rangle \right. \nonumber \\ 
&&+ \left. \cdots +
a^{(k)}_{^{N}C_k} |0_{1},\cdots,0_{N-k},1_{N-k+1},\cdots, 1_{N-1} \rangle\right\}.   
\end{eqnarray}     
It is to be noticed that the coefficients $a^{(r)}_{i}$ in (\ref{dgqubit}) are labeled such that $a^{(r)}_{i}$, $i=1,\,2,\ldots ^{N-1} C_r$ are associated with the states that have their $r$ spin-down qubits $\vert 1 \rangle$, permuted in the first $N-1$ positions, leaving the $N^{\rm th}$ position to $\vert 0\rangle$. The remaining coefficients $a^{(r)}_{i'}$, $i'~=~^{N-1} C_r +1,\,^{N-1} C_r +2,\ldots ^{N} C_r$ are associated with the states having their $N^{\rm th}$ position occupied by $\vert 1 \rangle$. Thus, $\vert \phi^{G}_0\rangle$ contains coefficients $a^{(r)}_{i}$, $i=1,\,2,\ldots ^{N-1} C_r$ whereas  $\vert \phi^{G}_1\rangle$ contains coefficients $a^{(r)}_{i'}$, $i'=^{N-1} C_r +1,\,^{N-1} C_r +2,\ldots ^{N} C_r$.led such that $a^{(r)}_{i}$, $i=1,\,2,\ldots ^{N-1} C_r$ are associated with the states that have their $r$ spin-down qubits $\vert 1 \rangle$, permuted in the first $N-1$ positions, leaving the $N^{\rm th}$ position to $\vert 0\rangle$. The remaining coefficients $a^{(r)}_{i'}$, $i'~=~^{N-1} C_r +1,\,^{N-1} C_r +2,\ldots ^{N} C_r$ are associated with the states having their $N^{\rm th}$ position occupied by $\vert 1 \rangle$. Thus, $\vert \phi^{G}_0\rangle$ contains coefficients $a^{(r)}_{i}$, $i=1,\,2,\ldots ^{N-1} C_r$ whereas  $\vert \phi^{G}_1\rangle$ contains coefficients $a^{(r)}_{i'}$, $i'=^{N-1} C_r +1,\,^{N-1} C_r +2,\ldots ^{N} C_r$.

From (\ref{dgphi01}), it can be readily seen that the $N-1$ qubit reduced density matrix $\rho^G_{1,2,\cdots, N-1}$ of the state $\vert D^{G}_{N-k,k}\rangle$ has the form 
\begin{equation}
\label{rhoG}
\rho^G_{1,2,\cdots, N-1}=|\phi^G_0\rangle\langle\phi^G_0|+|\phi^G_1\rangle\langle\phi^G_1|.
\end{equation}
where $\vert \phi^G_0\rangle$, $\phi^G_1\rangle$ are as given in (\ref{phig0}), (\ref{phig1}) respectively.   
If we demand that an $N$ qubit mixed state $\omega^{G}_{N}$ possesses the same $N-1$ qubit reduced state 
(\ref{rhoG}) then, there should exist an extended pure state $\vert\Omega^{G}_{NE}\rangle$ of $N$ qubits, appended with an environment $E$ in  such a way that  ${\rm Tr}_E[\vert\Omega^{G}_{NE}\rangle\langle \Omega^{G}_{NE}\vert]=\omega^{G}_{N}$ and   
\begin{equation}
\label{OmegaG}
\vert\Omega^G_{NE}\rangle=\vert \phi^G_0\rangle\vert E^G_0\rangle +\vert \phi^G_1\rangle
\vert E^G_1\rangle. 
\end{equation}
The states $\vert E^G_0\rangle, \vert E^G_1\rangle$ are comprised of the $N$th qubit  and the environment 
\begin{eqnarray*}
\vert E^G_0\rangle&=&\vert 0_N\rangle\,  \vert e^G_{00}\rangle+\vert 1_N\rangle\,  \vert e^G_{01}\rangle \\
\vert E^G_1\rangle&=&\vert 0_N\rangle\,  \vert e^G_{10}\rangle+\vert 1_N\rangle\,  \vert e^G_{11}\rangle 
\end{eqnarray*}
and they obey the orthonormality relations,
\begin{equation}
\label{onG}
\langle E^G_i\vert E^G_j\rangle=\delta_{i,j}.
\end{equation}     
The extended  pure state (\ref{OmegaG}) takes  the following form:  
\begin{eqnarray}
\label{Omega2G}
 \vert\Omega^G_{NE}\rangle=\vert \phi^G_0\rangle \vert 0_N\rangle\,\vert e^G_{00}\rangle +\vert \phi^G_0\rangle \vert
1_N\rangle\,\vert e^G_{01}\rangle+ \vert \phi^G_1\rangle \vert 0_N\rangle\,\vert e^G_{10}\rangle +\vert \phi^G_1\rangle \vert 1_N\rangle\,\vert 
e^G_{11}\rangle. 
\end{eqnarray} 
Having ascertained that $\vert D^{G}_{N-k,k}\rangle$ and $\omega^G_N$ possess a common reduced density matrix $\rho^G_{1,2,\cdots, N-1}$, we now impose that another $N-1$ qubit reduced density matrix $\rho_{2,3,\cdots, N}$ of $\vert D^{G}_{N-k,k}\rangle$ too is shared by $\omega^G_N$, or equivalently on $\vert\Omega^G_{NE}\rangle$. To verify the restrictions  on $\omega^G_N$, or on $\vert\Omega^G_{NE}\rangle$ due to this, we  compare the matrix elements of $\rho_{2,3,\cdots, N}={\rm Tr}_{1}\,[\vert D^G_{N-k,k}\rangle
\langle  D^G_{N-k,k}\rangle\vert]$ with that  obtained by tracing the 1st qubit, environment from $\vert\Omega^G_{NE}\rangle$.
\begin{enumerate} 
\item We first compare the following matrix elements:
\begin{eqnarray}
\label{one}
&&\langle 0_2,0_3,\cdots,0_{N-k-1},1_{N-k},\cdots 1_N\vert{\rm Tr}_{1,E}\,[\vert\Omega^G_{NE}\rangle
\langle  \Omega^G_{NE}\vert]\vert 
0_2,0_3,\cdots,0_{N-k-1},1_{N-k},\cdots 1_N\rangle
=|\alpha_k|^2 \vert a^{(k)}_{^{(N-1)}C_k}\vert^2 \langle e^G_{01}|e^G_{01}\rangle, \nonumber \\ 
&&\langle 0_2,0_3,\cdots,0_{N-k-1},1_{N-k},\cdots 1_N\vert{\rm Tr}_{1}\,[\vert D^G_{N-k,k}\rangle
\langle  D^G_{N-k,k}\rangle\vert]\vert 
0_2,0_3,\cdots,0_{N-k-1},1_{N-k},\cdots 1_N\rangle =0.  
\end{eqnarray}
Let us suppose that the coefficient $\vert a^{(k)}_{^{(N-1)}C_k}\vert$ is non-zero. We may then deduce that $\vert e^G_{01}\rangle \equiv 0$ (note that  $\alpha_k\neq 0$). 
The orthonormality relations $\langle E^G_0\vert E^G_0 \rangle=1$, $\langle E^G_0\vert E^G_1 \rangle=0$ would then lead to  
\begin{eqnarray}
\label{onG2}
\langle e^G_{00}\vert e^G_{00}\rangle=1, \ \ \langle e^G_{00}\vert e^G_{10}\rangle=0.  
\end{eqnarray}
\item Comparing yet another matrix element of $\rho_{2,3,\cdots, N}$ from both the pure states $\vert D^G_{N-k,k}\rangle$ and $\vert\Omega^G_{NE}\rangle$ 
(see Eqs.(\ref{dgqubit})--(\ref{phig1}), (\ref{Omega2G})), we obtain,  
\begin{eqnarray}
\label{2new}
&&\langle 0_2,0_3,\cdots,0_{N-k},1_{N-k+1},\cdots 1_N\vert{\rm Tr}_{1,E}\,[\vert\Omega^G_{NE}\rangle
\rangle\langle  \Omega^G\vert]\vert 
0_2,0_3\cdots,0_{N-k-1},1_{N-k},\cdots, 1_{N-1},0_N\rangle \nonumber \\
&=&\vert\alpha_k\vert^2   a^{(k)}_{^NC_k} \, a^{(k)*}_{^{N-1}C_k}\, \langle e^G_{11}|e^G_{00}\rangle,\nonumber \\
&&\langle 0_2,0_3,\cdots,0_{N-k},1_{N-k+1},\cdots 1_N\vert{\rm Tr}_{1}\,[\vert D^G_{N-k,k}\rangle
\rangle\langle  D^G_{N-k,k}\vert]\vert 
0_2,0_3,\cdots,0_{N-k-1},1_{N-k},\cdots, 1_{N-1},0_N\rangle  \nonumber \\
&=&\vert\alpha_k\vert^2   a^{(k)}_{^NC_k} \, a^{(k)*}_{^{N-1}C_k}.
\end{eqnarray}
As  $\vert a^{(k)}_{^{(N-1)}C_k}\vert\neq 0$, and assuming that $a^{(k)}_{^NC_k}\neq 0$, (\ref{2new}) results in the condition,  
\begin{equation}
\langle e^G_{11}|e^G_{00}\rangle=1.
\end{equation}  
This, together with the relation $\langle e^G_{00}\vert e^G_{00}\rangle=1$ yields  
\begin{equation}
\vert e^G_{11}\rangle\equiv \vert e^G_{00}\rangle. 
\end{equation}
It is not difficult to see that 
\begin{eqnarray}
\langle E^G_{1}\vert E^G_{1}\rangle=1&& 
\Rightarrow  \langle e^G_{10}\vert e^G_{10} \rangle+\langle e^G_{11}\vert e^G_{11} \rangle=1 \nonumber \\
&&\Rightarrow \langle e^G_{10}\vert e^G_{10} \rangle+\langle e^G_{00}\vert e^G_{00} \rangle=1 \nonumber \\ 
&& \Rightarrow  \langle e^G_{10}\vert e^G_{10}\rangle=0 \ \ ({\mbox{as}}\ \ \langle e^G_{00}\vert e^G_{00} \rangle=1) \nonumber \\
& &\Rightarrow \vert e^G_{10}\rangle=0. 
\end{eqnarray}
Finally, on substituting $\vert e^G_{01}\rangle\equiv \vert e^G_{10}\rangle=0$, $\vert e^G_{00}\rangle\equiv \vert e^G_{11}\rangle=1$ in (\ref{Omega2G}), we get the explicit form of the state $\vert\Omega^G_{NE}\rangle$ as,
\begin{eqnarray}
\label{final}
|\Omega^G_{NE}\rangle&=& \vert \phi^G_0\rangle \vert 0_N\rangle\,\vert e^G_{00}\rangle+\vert \phi^G_1\rangle \vert 1_N\rangle\,\vert e^G_{00}\rangle \nonumber \\
&=&\vert D^G_{N-k,\,k}\rangle\, \vert e^G_{00}\rangle
\end{eqnarray}
with $\vert D^{G}_{N-k,k}\rangle=\vert \phi^{G}_0\rangle \vert 0\rangle_N + 
\vert \phi_1^{G}\rangle \vert 1\rangle_N$ (see (\ref{dgphi01})).
\end{enumerate}

We thus come to the  conclusion, by employing only {\em two} of the $N-1$ qubit reduced density matrices, that the generalized  states $\vert D^G_{N-k,\,k}\rangle$ are uniquely determined by their $N-1$ party marginals.  

It is important to note here that while the above result perfectly holds good for the class of states $\{|D^G_{N-k,\,k}\rangle\}$ of Eq.~(\ref{dgqubit}) when all the coefficients $\alpha_r,\ a^{(r)}_{i}, i=0,1,2,\ldots,^N C_r, r=0,1,\ldots, k$ are non-zero, it is valid if at least the coefficients  $a^{(k)}_{^N\, C_k}$ and $a^{(k)}_{^{N-1}\, C_k}$, in  Eq.~(\ref{dgqubit}) are non-zero (because the matrix elements of $N-1$ qubit reduced states given in Eqs.~(\ref{one}), (\ref{2new}) vanish if the coefficients  $a^{(k)}_{^N\, C_k}$ and $a^{(k)}_{^{N-1}\, C_k}$ are zero and therefore the inferences drawn from the Eqs.~(\ref{one}), (\ref{2new}) do not hold good in such cases).      
Based on  the possibility of different choices of $N-1$ qubit reduced matrix elements to arrive at the same result (\ref{final}), we arrive at the conclusion that  a unique specification of the generalized class of states $\{{\cal D}^G_{N-k,\,k}\}$ -- using only {\em two} of their $N-1$ qubit marginals -- is possible provided both the conditions given below are satisfied:

\begin{itemize}
\item among the set of coefficients $\{ a^{(k)}_{s}, s=1,2,\ldots ^{N-1} C_k\}$ (see Eq.~(\ref{dgqubit})) for a given k=0,1,2\ldots,[N/2],  {\em at least} one coefficient -- that {\em contains}   $\vert 0\rangle$ in the first qubit position -- is non zero.
\item among the remaining coefficients in $\{ a^{(k)}_{s'} \}$, $s'=^{N-1} C_k+1,\, ^{N-1} C_k+2 \ldots ^{N} C_k$ at least one coefficient -- with its first qubit position {\em  occupied} by $\vert 0\rangle$  -- is non-vanishing. 
\end{itemize}  
Excluding the class of states not obeying the above two conditions, all other states
belonging to the generalized class of states $\{{\cal D}^G
_{N-k, k}\}$ belong uniquely to their $N - 1$ party marginals. It is illuminating to note that in spite of the generality
of this class of non-symmetric states, only {\em two} of the  $N - 1$ qubit marginals suffice
for their unique determination.

\section{Geometric measure of entanglement}

Quantification of multiparty entanglement forms one of the central themes underlying  quantum information theory. Several entanglement measures have been proposed in this context~\cite{review1,review2}, though they suffer because of the optimization involved in their evaluation. Natural strategy towards this end has been to restrict to certain class of states  obeying specific symmetries in order to carry out such optimization procedures.

Let us consider the widely employed geometric measure of entanglement~\cite{Shimony, Wei} associated with a multiparty pure state $\vert \psi\rangle$: 
\begin{equation}
\label{gm} 
E_G(\vert\psi\rangle)=
%1-\Lambda^2_{\rm max},  \ \ {\Lambda_{\rm max}}=
1-\max_{\{\vert\epsilon^{\rm prod}\rangle\}}\vert\langle \epsilon^{\rm prod}\vert \psi\rangle\vert^2 
\end{equation}            
  where $\{\vert\epsilon^{\rm prod}\rangle\}$ is the set of {\em all} pure separable (product) states.  
Another equivalent quantification of the geometric measure~\cite{markham2} is given by, 
\begin{equation}
\label{loggm}
{\cal E}_G(\psi)=-{\rm log}_2\left[ \max_{\{\vert\epsilon^{\rm prod}\rangle\}}\vert\langle \epsilon^{\rm prod}\vert \psi\rangle\vert^2\right].
\end{equation}

For $N$ qubit GHZ states $\vert{\rm GHZ}\rangle=\frac{1}{\sqrt{2}}[\vert 0,0,\ldots,0\rangle+\vert 1,1,\ldots, 1\rangle]$ the geometric measure 
$E_G(\vert{\rm GHZ}\rangle)=\frac{1}{2}$  and the logarithmic geometric measure ${\cal E}_G(\vert{\rm GHZ}\rangle)=1$ --  independent of the number of qubits~\cite{Wei}. The geometric measure for the Dicke states (\ref{Dicke}) is found to be~\cite{Wei} 
\begin{equation}
\label{gmdicke}
E_G\left(\left\vert\frac{N}{2},\frac{N}{2}-l\right\rangle\right)=1-^N C_l\, \left(\frac{l}{N}\right)^l\, \left(\frac{N-l}{N}\right)^{N-l},
\end{equation}  
which takes its maximum value when $l$ is closest to $N/2$.

The optimization procedure in evaluating the geometric measure (\ref{gm}) is  non-trivial in the case of general multiparty states. In this connection, a great deal of attention has been drawn to address the question: "Is the closest separable state of an arbitrary symmetric multiparty state $\vert \psi_{\rm sym}\rangle$ itself a symmetric product state?"~\cite{Wei, clsep,clsep1,clsep2}. It is only very recently~\cite{ogunhe} that it has been established that the optimal state (closest separable state) maximizing the geometric measure $E_G(\vert\psi\rangle)$ of (\ref{gm}) is necessarily symmetric for three or more party states obeying exchange symmetry~\cite{clsep2}. This  identification amounts to considerable simplification in the evaluation of the geometric measure of entanglement of pure permutation symmetric multiqubit  states. 

The closest product state associated with the Dicke states (\ref{Dicke}) leading to optimization of the geometric measure is found to be~\cite{clsep}
\begin{equation}
\label{clsdicke}
\vert \epsilon^{\rm prod}_{N, l}\rangle=\left(\sqrt{\frac{N-l}{N}}\,\vert 0\rangle+\sqrt{\frac{l}{N}}\, \vert 1 \rangle \right)^{\otimes N}.
\end{equation}
and this yields the amount of geometric entanglement given by (\ref{gmdicke}).

The MR~(\ref{Maj}) of symmetric multiqubit states is very useful to obtain a simplified structure for 
 the geometric measure of entanglement. Substituting Eq.~(\ref{Maj}) into 
 Eq.~(\ref{gm}) and considering that the maximization is only
required over the set of symmetric separable states $\vert\epsilon,\epsilon,\ldots, \epsilon\rangle$ leads to 
\begin{equation}
\label{gm2} 
E_G(\vert\Psi_{\rm sym}\rangle)=
1-{\cal N}^2\, N!^2\, \max_{\{\vert\epsilon\rangle\}}\,\prod_{i=1}^{N}\vert\langle \epsilon\vert \epsilon_i\rangle\vert^2 
\end{equation}

We recall that MR maps every permuatation symmetric state $\vert \Psi_{\rm sym} \rangle$ to $N$ points  on the unit sphere ( these points are referred to as the Majorana points (MP) ~\cite{markham2}). The point on the Majorana sphere corresponding to the state $\vert \epsilon \rangle$ which optimizes the geometric measure $E_G(\vert\Psi_{\rm sym}\rangle)$ 
in (\ref{gm2}) is called the closest product point(CPP)~\cite{markham2}.
 
Aulbach et.al.,~\cite{markham2} evaluated the geometric measure of entanglement for some well-known two and three qubit symmetric states by  making  use of the MR as follows:  Any identical local unitary operation on each spinor of the two qubit state 
$\vert \Psi_{\rm sym} \rangle={\cal N}\ [\vert\epsilon_1,\epsilon_2\rangle+\vert\epsilon_2,\epsilon_1\rangle]$ is equivalent to a rotation of MPs about a common axis on the Majorana sphere. Making use of identical local unitary transformation, a given distribution of two MPs can be rotated on the Majorana sphere in such a way that both the points lie in the positive hemisphere. In other words, the two spinors constituting a two-qubit symmetric state can be rotated so as to obtain $\vert \epsilon'_1 \rangle= \vert 0\rangle$ and $\vert \epsilon'_2\rangle=\cos \frac{\theta}{2} \vert 0 \rangle+\sin \frac{\theta}{2} \vert 1 \rangle,$  $0\leq \theta\leq \pi$. The closest separable state of the two qubit symmetric state constituted by the spinors $\vert \epsilon'_1\rangle,\ \vert \epsilon'_2\rangle$  is identified as  $\vert \epsilon \rangle=\cos \frac{\theta}{4} \vert 0 \rangle+\sin \frac{\theta}{4} \vert 1 \rangle$. (i) When $\theta=0$, one gets the separable state $\vert 0,0\rangle$ and the  geometric measure (see (\ref{gm2}) vanishes.  (ii) Choosing $\theta=\pi$, one obtains the Bell state $\vert \Psi^+ \rangle=\frac{1}{\sqrt 2}
\left( \vert 0,1 \rangle+\vert 1,0 \rangle \right)\rangle$. The corresponding closest separable state is given by, 
$\vert \epsilon \rangle=\frac{1}{\sqrt{2}} [\vert 0 \rangle+\vert 1 \rangle]$ (note that the entire set of  states 
$\vert 0 \rangle+e^{i\phi}\vert 1 \rangle$, which form a continuous ring around the equator on the Majorana sphere correspond to closest separable states of  the Bell state $\vert \Psi^+ \rangle$) and the geometric measure of entanglement is given by 
$E_G(\vert \Psi^+ \rangle)=\frac{1}{2}$ (the logarithmic geometric measure ${\cal E}_G(\vert \Psi^+ \rangle)=1$).

The Majorana spinors constituting  the $3$ qubit GHZ state are given by (see Table~1)   
$\vert \epsilon_1\rangle= \frac{1}{\sqrt{2}}(\vert 0\rangle+
\vert 1\rangle)$, $\vert \epsilon_2\rangle= \frac{1}{\sqrt{2}}(\vert 0\rangle+e^{2i\pi/3}
\vert 1\rangle)$, $\vert \epsilon_3\rangle= \frac{1}{\sqrt{2}}(\vert 0\rangle+e^{4i\pi/3}
\vert 1\rangle)$  (upto an overall phase); the CPP states are identified~\cite{markham2} to be  $\vert 0 \rangle$, $\vert 1 \rangle$. 
Thus, the geometric measure of entanglement is readily evaluated to obtain $E_G(\vert{\rm GHZ}\rangle)=\frac{1}{2}$ and the associated logarithmic measure (see (\ref{loggm}) ${\cal E}_G(\vert{\rm GHZ}\rangle)=1.$ 

The Majorana spinors associated with the 3-qubit W state $\vert{\rm W}\rangle=\frac{1}{\sqrt{3}}[\vert 0,0,1\rangle+\vert 0,1,0\rangle+\vert 1,0,0\rangle]$ are given by (see Table~1)  $\vert \epsilon_1 \rangle=\vert 0\rangle,\, \vert \epsilon_2 \rangle=\epsilon_3\vert=\vert 1\rangle$ and the CPP state is given by $\vert\epsilon\rangle=\sqrt{\frac{2}{3}}\,\vert 0\rangle+\sqrt{\frac{1}{3}}\, \vert 1 \rangle$.  A continuous ring given by $\sqrt{\frac{2}{3}}\,\vert 0\rangle+e^{i\phi}\sqrt{\frac{1}{3}}\, \vert 1 \rangle$ forms a set of CPP states~\cite{markham2} of the three qubit W state, entanglement of which is therefore found to be $E_G(\vert W\rangle)=\frac{5}{9}.$

An approach to evaluate the geometric measure of entanglement of symmetric multiqubit states (by identifying CPP states geometrically)  and to identify maximally entangled symmetric states -- by   exploiting the symmetries of the MP distribution --  has been discussed in detail in Ref.~\cite{markham2}. Further, an improved asymptotic trend  of the geometric measure of maximally entangled symmetric states (compared to that of highly entangled Dicke states (\ref{Dicke}) with $l=[N/2]$) has also been identified~\cite{gebastin}.

 Instead of identifying  CPP states geometrically~\cite{markham2},  
we propose to use the  collective representation of the symmetric states as follows:  
The set of all symmetric $N$-qubit product states $\{\vert \phi^{\rm prod}_{\rm sym}\rangle=\vert \epsilon,  \epsilon, \ldots , \epsilon\rangle\}$  can be collectively represented as  the  spin coherent states~\cite{Arechi} i.e., 
\begin{eqnarray} 
  \label{scs}
   \vert \phi^{\rm prod}_{\rm sym}\rangle&\equiv&\vert \alpha,\beta\rangle=e^{\tau\, J_+-\tau^*J_-}\, 
   \left\vert \frac{N}{2}, -\frac{N}{2}\right\rangle \nonumber \\ 
   &=& \sum_{r=0}^{N}\, \sqrt{^N C_{r}}\, \left(\cos\frac{\beta}{2}\right)^{r}\, \left(\sin\frac{\beta}{2}\right)^{N-r}\,  e^{-i\,(N-r)\,\alpha}\  \left\vert \frac{N}{2}, \frac{N}{2}-r\right\rangle.  
\end{eqnarray} 
%where $\left\vert \frac{N}{2}, l-\frac{N}{2}\right\rangle$  denote the symmetric Dicke states (see Eq.~(\ref{Dicke});
where $\tau=\frac{\beta}{2}\, e^{i\alpha}$, 
$0\leq \alpha\leq 2\pi$, $0\leq \beta\leq \pi$;  $J_\pm=\sum_{i=1}^{N}\, \sigma_{i\pm}$ are the collective spin ladder operators, and $\sigma_{i\pm}=\sigma_{ix}\pm\sigma_{iy}$ denote the Pauli operators of the $i$th qubit.    
Employing this collective spin coherent state representation of symmetric product states, we may express the geometric measure of entanglement for permutation symmetric states (\ref{sympure1}) as, 
\begin{eqnarray}
\label{gemsym}
E_G(\vert\Psi_{sym}\rangle)&=&1-\max_{\{\alpha,\beta\}}\, F(\alpha,\beta),\\
F(\alpha,\beta)&=&\left\vert\sum_l\, c_l\, \right\langle \alpha,\beta\left\vert \frac{N}{2}, \,\left.\frac{N}{2}-l\right\rangle\right\vert^2.
\end{eqnarray}
Here, the optimization is done over the set of angles $\alpha,\beta$ of the collective spin coherent states (\ref{scs}) -- thus offering a novel method for evaluating the geometric entanglement of symmetric states.

For example, let us consider the Dicke states Eq.~(\ref{Dicke}): We  simplify the maximum value of $F(\alpha,\beta)$ as follows:  
\begin{eqnarray}
\label{scopt}
\max_{\{\alpha,\beta\}}F(\alpha,\beta)&=&
\max_{\{\alpha,\beta\}}\left\vert\langle \alpha,\beta\vert \frac{N}{2}, \,\frac{N}{2}-l\rangle\right\vert^2\nonumber \\
&=&\max_{\{\beta\}}\,F_{N,l}(\beta)=\max_{\{\beta\}}\, \left[^N C_l\,  \left(\cos\frac{\beta}{2}\right)^{2l}\, 
\left(\sin\frac{\beta}{2}\right)^{2(N-l)} \right].% \nonumber \\
%&=&^N\, C_l\, \left(\cos\frac{\beta}{2}\right)^{2l}\, \left(\cos\frac{\beta_M}{2}\right)^{2l}\, 
%\left(\sin\frac{\beta_M}{2}\right)^{2(N-l)},
\end{eqnarray}  
In order to obtain the maximum value of the function  $F_{N,l}(\beta)$ in Eq.~(\ref{scopt}) we consider $\log\, [F_{N,l}(\beta)$ and set 
$\left.\frac{d\log[F_{N,l}(\beta)]}{d\beta}\right\vert_{\beta_M}=0$. We obtain, 
\begin{eqnarray}
\label{scopt2} 
 \left.\frac{d\log[F_{N,l}(\beta)]}{d\beta}\right\vert_{\beta_M}&=&-l\, \tan\frac{\beta_M}{2}+(N-l)\, \cot\frac{\beta_M}{2}=0 \nonumber \\
 \Rightarrow &&  \tan\frac{\beta_M}{2}=\pm\sqrt{\frac{N-l}{l}}.  
\end{eqnarray}
We thus obtain, 
\begin{equation}
F_{N,l}(\beta_M)]=^N C_l\, \left(\frac{l}{N}\right)^l\,\left(\frac{N-l}{N}\right)^{N-l}
\end{equation}
which readily leads to the geometric measure of entanglement (\ref{gmdicke}) of Dicke states and also to the identification of their closest product states (\ref{clsdicke}). 

For the $N$-qubit GHZ state 
\begin{equation}
\label{ghzn}
\vert {\rm GHZ}\rangle=\frac{1}{\sqrt{2}}\left[\left\vert \frac{N}{2},\frac{N}{2}\right\rangle+\left\vert \frac{N}{2},-\frac{N}{2}\right\rangle\right] 
\end{equation}
we obtain 
\begin{equation}
\max_{\{\alpha,\beta\}}F_{\rm GHZ}(\alpha,\beta)=\max_{\{\alpha,\beta\}}\left\vert\langle \alpha,\beta\vert {\rm GHZ}\rangle\right\vert^2
=\frac{1}{2}\left[\left(\sin\frac{\beta}{2}\right)^{2N}+\left(\cos\frac{\beta}{2}\right)^{2N}+2\left(\cos\frac{\beta}{2}\sin\frac{\beta}{2}\right)^{N}\cos(N\alpha)\right]. 
\end{equation} 
Optimal value of the function $F_{\rm GHZ}(\alpha,\beta)$ is obtained for $\alpha_M={\rm arbitrary},\ \beta_M=0$ leading to 
$F_{\rm GHZ}(\alpha_M,\beta_M)=\frac{1}{2}$ -- in agreement with the earlier result~\cite{Wei}. 
Note that the geometric measure of entanglement of GHZ state is less than that of the Dicke state (\ref{Dicke})) with $l=[N/2]$,  indicating different hierarchies of multiparticle entanglement (depending on the nature of the measure employed). 
It would be interesting to explore expansions of $N$-particle symmetric states in terms of $p$-particle constituents~\cite{Ruskai} (in particular, those with $p=2$ are called  geminal expansions) in order to  recognize genuine multiparticle entanglement in a physically significant manner.

\section{Summary}

This article presents a detailed description of the Majorana geometrical representation of  symmetric multiqubit states. With the help of the MR, the SLOCC entanglement classification of pure symmetric states in terms of the number and arrangement of the distinct Majorana spinors constituting them is elucidated. Further,  uniqueness of the whole pure symmetric $N$-qubit states belonging to the two distinct spinor family -- and also, another related class of non-symmetric states -- to their $N-1$ qubit reduced density matrices is established (by employing only {\em two} of the reduced states). Quantification of multiqubit entanglement of permutation symmetric states in terms of  geometric measure of entanglement (where the MR has been employed extensively) is detailed.

\section*{Acknowledgements}
Sudha gratefully acknowledges the support of
D.C.Pavate Foundation for the award of Pavate Memorial Visiting Fellowship. She is also thankful to the
local hospitality and facilities provided at Sidney Sussex
College, Cambridge, UK.

\end{document}